\documentclass[aps,twocolumn,showpacs,preprintnumbers,amsmath,amssymb,nofootinbib,superscriptaddress,showkeys]{revtex4}

\usepackage{epsfig}
\usepackage{graphicx}
\def\negate{\hspace{-7.0pt}/ \,}

\begin{document}

\title{Renormalization of chiral two-pion exchange NN interactions
with $\Delta$-excitations: Central Phases and the Deuteron.}
\author{M. Pav\'on
  Valderrama}\email{m.pavon.valderrama@fz-juelich.de}
  \affiliation{Institut f\"ur Kernphysik and J\"ulich Center for Hadron 
    Physics, Forschungszentrum J\"ulich, 52425 J\"ulich, Germany} 
\author{E. Ruiz
  Arriola}\email{earriola@ugr.es} \affiliation{Departamento de
  F\'{\i}sica At\'omica, Molecular y Nuclear, Universidad de Granada,
  E-18071 Granada, Spain.}
\date{\today}

\begin{abstract} 
\rule{0ex}{3ex} The renormalization of the chiral np interaction with
$\Delta$-excitations in intermediate states is considered at next-to-leading 
order (NLO) and next-to-next-to-leading order (N2LO) for central waves.
The inclusion of the $\Delta$-excitation as an explicit degree of freedom
improves the convergence properties of the effective field theory 
results for np scattering with respect to $\Delta$-less theory,
and allows the existence of a deuteron bound state in the infinite 
cut-off limit.
The $^1S_0$ singlet and $^3S_1$ triplet phase shifts reproduce data 
for $p \sim m_\pi$.
The role of spectral function regularization is also discussed.

\end{abstract}
\pacs{03.65.Nk,11.10.Gh,13.75.Cs,21.30.Fe,21.45.+v} \keywords{NN
interaction, One and Two Pion Exchange,
$\Delta$-excitations. Renormalization.}

\maketitle

%\tableofcontents

%=========================================

\section{Introduction} 

The nucleon-nucleon (NN) interaction problem has a well deserved
reputation of being a difficult one (for a review see
e.g.~\cite{Machleidt:2001rw}).  One obvious reason may be found in the
theoretical and experimental inaccessibility of the shortest distance
interaction region relevant to Nuclear Physics. This lack of
fundamental knowledge might be remedied in the future by {\it ab
  initio} lattice calculations, for which some incipient results exist
already~\cite{Beane:2006mx,Ishii:2006ec}. However, the current
situation does not prevent from addressing many facets of NN
interactions, particularly long distance features, provided short
distance insensitivity is guaranteed. Despite the great efforts during
the years based on successful
phenomenology~\cite{Machleidt:1987hj,Stoks:1993tb,Stoks:1994wp,Machleidt:2000ge},
only during the last decade has the parentage to the underlying QCD
been made more explicit after the
proposal~\cite{Weinberg:1990rz,Ordonez:1995rz} to use Effective Field
Theory (EFT) methods and extensive use of Chiral Symmetry (for
comprehensive reviews see
e.g. Ref.~\cite{Bedaque:2002mn,Epelbaum:2005pn,Machleidt:2005uz}).
EFT starts out with the most general Lagrangean in terms of the
relevant degrees of freedom compatible with chiral symmetry,
dimensional analysis and perturbative renormalization, organized by a
prescribed power counting based on assuming a large scale suppression
on the parameters $ 4 \pi f_\pi \sim M_N \sim 1 {\rm GeV}$. Thus, by
construction, EFT complies to the expectation of short distance
insensitivity, provided that enough counter-terms encoding the unknown
short distance physics are added in perturbation theory to a finite
order~\cite{Ordonez:1995rz,Kaiser:1997mw,Kaiser:1998wa,Friar:1999sj,Rentmeester:1999vw,Kaplan:1998tg,Kaplan:1998we,Beane:2001bc,Frederico:1999ps}.
However, the physics of bound states, such as the deuteron, is
genuinely non-perturbative and an infinite resummation of diagrams
proves necessary (see however~\cite{Beane:2008bt} for a renewed
perturbative set up). There arises the problem of non-perturbative
renormalization and/or modifications of the original power counting
for which no universally accepted scheme has been found yet, mainly
because the requirements may vary.  Thus, although the EFT method does
comply to model independence it is far from trivial to achieve
regulator independence and a converging pattern dictated by a
reasonable power counting within a non-perturbative set up.

The NN renormalization problem may be cast efficiently in the
traditional language of potentials and the corresponding Schr\"odinger
equation. Besides phrasing the NN interaction into the familiar and
more intuitive non-relativistic quantum mechanical framework, this
procedure has also the advantage that the perturbative determination
of the potential complies to the desired short distance insensitivity
for the potential
itself~\cite{Ordonez:1995rz,Kaiser:1997mw,Kaiser:1998wa,Friar:1999sj,Rentmeester:1999vw,Entem:2002sf}. The
unconventional feature is that chiral expansions necessarily involve
singular potentials at short distances, i.e. $r^2 |V(r)| \to \infty $
for $r\to 0$. Indeed, in the limit $r \ll 1/m_\pi $ (or equivalently
large momenta) pion mass effects are irrelevant in loop integrals and
hence at some fixed order of the expansion one has for the coupled
channel potential~\footnote{The only exception is the singlet
channel-One-Pion-Exchange (OPE) case which behaves as $\sim m_\pi^2/
f_\pi^2 r$, see below.}
\begin{eqnarray}
 V(r) \to \frac{M_N {\bf c}_{2n+m+r}}{(4 \pi f_\pi)^{2 n} M_N^m \Delta^r}
 \frac{1}{r^{2 n+m + r}} \, ,
\label{eq:pot-sing}
\end{eqnarray}
where $n$, $m$ and $r$ are nonnegative ($\ge 0$) integers, $\Delta$ the 
N$\Delta$ splitting, 
and ${\bf c}_{k}$, with $k = 2n + m + r$, is a dimensionless matrix of van der
Waals coefficients in coupled channel space. The dimensional argument
is reproduced by loop calculations in the Weinberg dimensional power
counting~\cite{Ordonez:1995rz,Kaiser:1997mw,Kaiser:1998wa,Friar:1999sj,Rentmeester:1999vw,Entem:2001cg,Entem:2002sf,Entem:2003ft}
and is scheme independent.  It is thus conceivable that much of our
understanding on the physics deduced from chiral potentials might be
related to a proper interpretation of these highly singular
potentials, which degree of singularity increases with the order of
the expansion. This of course raises the question about in what sense
higher order potentials are smaller. One obvious way to ensure this
desired smallness is by keeping a finite and sufficiently moderate
cut-off~\cite{Ordonez:1995rz,Entem:2001cg,Entem:2003ft,Epelbaum:1999dj,Epelbaum:2003gr,Epelbaum:2003xx,Epelbaum:2004fk},
so the singularity of the potential is not probed, effectively
recovering the power counting expectations for the size of higher
order contributions.  The disadvantage is that results are strongly
cut-off dependent for scales about $r_c \sim 0.5 - 1.0\,{\rm fm}$,
similar to the ones probed in NN
scattering~\cite{PavonValderrama:2005gu,Valderrama:2005wv,PavonValderrama:2005uj,Nogga:2005hy}.
However, this is not the only possibility.  Explicit
computations~\cite{PavonValderrama:2005gu,Valderrama:2005wv,PavonValderrama:2005uj}
show that, when the cut-off is removed, the higher order contributions
will continue to generate only small changes in the amplitudes under
specific circumstances which can be determined {\it a priori}. As
emphasized in previous
works~\cite{PavonValderrama:2005gu,Valderrama:2005wv,PavonValderrama:2005uj}
renormalization is the most natural tool provided 1) we expect the
potential to be realistic at long distances and 2) we want short
distance details to be inessential in the description. As discussed
above, this is precisely the situation we face most often in Nuclear
Physics. A surprising and intriguing feature is that knowledge on the
attractive or repulsive character of the singularity, i.e.  the sign
of the eigenvalues of ${\bf c}_k$ in Eq.~(\ref{eq:pot-sing}), turns
out to be crucial to successfully achieve this program and ultimately
depends on the particular scheme or power counting used to compute the
potential. 

The singularities of chiral potentials may be disconcerting~\footnote{
  Early treatments can be found in \cite{Case:1950an} (see
  Ref.~\cite{Frank:1971xx} for an early review).}, but they can be
handled in a way that do not differ much from the standard treatment
of well-behaved regular potentials which one usually encounters in
nuclear physics~\cite{RuizArriola:2007wm}. Renormalization is the
mathematical implementation of the appealing physical requirement of
short distance insensitivity and hence a convenient tool to search for
typical long distance model- and regulator independent results. In a
non-perturbative setup such as the NN problem, renormalization imposes
rather tight constraints on the interplay between the unknown short
distance physics and the perturbatively computable long distance
interactions~\cite{PavonValderrama:2005gu,Valderrama:2005wv,PavonValderrama:2005uj}.
This viewpoint provides useful insights and it is within such a
framework that we envisage a systematic and model independent
description of the NN force based on chiral interactions.  In this
regard it is amazing to note how the sophisticated machinery of
perturbative renormalization in Quantum Field Theory used to compute
the chiral potentials has not been so extensively developed when the
inevitable non-perturbative physics must be incorporated; quite often
the renormalization process is implemented by trial and error methods
by adding counter-terms suggested by the a priori power counting in
momentum space. However, detailed analyses in coordinate and momentum
space~\cite{PavonValderrama:2005gu,Valderrama:2005wv,PavonValderrama:2005uj,PavonValderrama:2007nu,Beane:2000wh,Long:2007vp}
show that the allowed structure of counter-terms can be anticipated on
purely analytical grounds and cannot be chosen independently on the
long distance potential (see also Refs.~\cite{Yang:2007hb,Yang:2009kx}
for numerical work). Specifically, for a channel-subspace with good
total angular momentum the number of counterterms is $n(n+1)/2$ with
$n$ the number of negative eigenvalues of the van der Waals matrix
${\bf c}_k$ in Eq.~(\ref{eq:pot-sing}).  In fact, the much simpler
coordinate space renormalization has been shown to be fully equivalent
to the popular momentum space treatment for purely contact
theories~\cite{vanKolck:1998bw} and theories containing additional
long range physics~\cite{Entem:2007jg,Valderrama:2007ja}.  On the
light of these latter studies the smallness of increasingly singular
potentials such as Eq.~(\ref{eq:pot-sing}) is triggered quite
naturally by choosing the regular solution of the Schr\"odinger
equation; the wave function behaves as $u_p(r) \sim ( 4\pi f_\pi r
)^{\frac{2n+m+r}4} $ (modulo prefactors depending on the attractive or
repulsive character of the potential). Thus, as shown in
Ref.~\cite{Valderrama:2007ja}, when the short distance cut-off $r_c$
approaches a {\it fixed scale}, $\sim 1/(4 \pi f_\pi) =0.2 {\rm fm} $,
an increasing ${\cal O} ( r_c^{n+m/2+r/2+1})$ insensitivity of phase
shifts and deuteron properties is guaranteed as the power of the
singularity of the potential increases. Indeed, calculations with
chiral Two Pion Exchange (TPE) potentials reproducing low energy NN
data display this insensitivity for reasonable scales of $r_c \sim 0.5
{\rm fm}$~\cite{PavonValderrama:2005gu,Valderrama:2005wv} which
correspond to the shortest wavelength probed by NN scattering in the
elastic region. Within such a scenario, the discussion on whether one
should remove the cut-off or
not~\cite{PavonValderrama:2005gu,Valderrama:2005wv,Nogga:2005hy,Epelbaum:2006pt,Birse:2005um}
would become less relevant as the order of the chiral expansion is
increased in the computations. This requires of course that the same
renormalization conditions are implemented, in order for the
computations to represent an specific physical situation, and that the
cut-off lies in the stability region $r_c \sim 0.5 {\rm fm}$, which in
turn means that renormalization has effectively been achieved.

However, despite all these findings, the question on how a sensible
hierarchy for NN interactions should be organized has been left
open. Although we know {\it whether} and, in positive case, {\it how}
this can be made compatible with the desired short distance
insensitivity~\cite{PavonValderrama:2005gu,Valderrama:2005wv,PavonValderrama:2005uj},
not all chiral potentials based on any given power counting are
necessarily eligible. The Weinberg counting based in a heavy baryon
approach at leading order (LO)~\cite{Epelbaum:2005pn} for $^1S_0$ and
$^3S_1-{}^3D_1$ states turns out to be
renormalizable~\cite{PavonValderrama:2005gu} due to an
attractive-repulsive character of the coupled channel eigen potentials
at short distances. There is at present no logical need why this
ought to be so, for the simple reason that power counting does not
anticipate the sign of the interaction at short distances. 
In fact when one goes to next-to-leading order (NLO) the
short distance $1/r^5$ singular repulsive character of the potential
makes the deuteron unbound~\cite{Valderrama:2005wv} because the
interaction becomes singular and repulsive. Finally,
next-to-next-to-leading order (NNLO) potentials diverge as $-1/r^6$
and are, again, compatible with Weinberg counting in the
deuteron~\cite{Valderrama:2005wv}, in this case because the
interaction is singular and attractive. Further inconsistencies
between Weinberg's power counting and renormalization have been
reported in Refs.~\cite{Nogga:2005hy,Entem:2007jg}. The previous
examples show clearly that the requirement of renormalizability can be
in open conflict with the idea of a convergent pattern based on a
preconceived chiral expansion and the mere dimensional power counting
of the original proposal~\cite{Weinberg:1990rz}. In particular, the
assumption that the NN potential cannot be completely determined at
arbitrary short distances is incompatible with a fully repulsive short
distance singularity because finiteness implies that no counter-term is
allowed.~\footnote{In the coupled channel case that means {\it all}
eigen-channels being repulsive at short distances.}

In this regard one should note an interesting analogy between the NN
interaction in the chiral quark model (for a review see
e.g.~\cite{Entem:2000mq,Bartz:2000vm,Valcarce:2005em}) and the van der Waals 
molecular interactions in the Born-Oppenheimer
approximation~\cite{Valderrama:2005wv}.  For non-relativistic
constituent quarks the NN interaction is provided by the convoluted
OPE quark-quark potential in the direct channel since quark exchange
and finite nucleon size effects can be neglected for distances larger
than the nucleon size.  Second order perturbation theory in OPE among
quarks generates TPE between nucleons yielding
\begin{eqnarray}
V_{NN} &=& \langle NN | V_{\rm OPE} | NN \rangle \nonumber \\ 
&+& \sum_{HH' \neq NN
}\frac{ |\langle NN | V_{\rm OPE} | HH' \rangle |^2}{ E_{NN}-E_{HH'}
}+ \dots
\label{eq:oppenheimer} 
\end{eqnarray} 
for the NN potential, where $| NN \rangle$ represents a two-nucleon state
and $| HH' \rangle$ an arbitrary intermediate state. 
In Regge theory one has the relation $M_\Delta^2
- M_N^2 = m_\rho^2 - m_\pi^2$~\cite{Ademollo:1969nx}, $m_{\rho}$ being
the rho mass, which in the chiral limit and for $M_N \sim N_c m_\rho/ 2 $ 
implies a $N\Delta$ splitting 
$\Delta \sim m_\rho^2 / 2 M_N \sim m_\rho /N_c = 256 {\rm  MeV}$ 
for $N_c =3 $ and suggests a scale numerically comparable to
twice the pion mass, which actually vanishes for $N_c \to \infty$. 
Thus, it makes sense to consider an EFT where the $\Delta N$
splitting is regarded as a small
parameter~\cite{Jenkins:1990jv,Hemmert:1997ye}.
Actually, when $H H'=N \Delta$ and $H
H'=\Delta \Delta$, Eq.~(\ref{eq:oppenheimer}) resembles the result
found using the Feynman graph
technique~\cite{Kaiser:1998wa}. Moreover, the second order
perturbative character suggests that the potential becomes attractive,
since $E_{NN}-E_{N\Delta}= -\Delta$ and $E_{NN}-E_{\Delta\Delta}=
-2\Delta$. At short distances the matrix elements scale as $\langle NN
| V_{\rm OPE} | HH' \rangle \sim g_A^2 /(f_\pi^2 r^3)$ and hence the
potential becomes singular $\sim -g_A^4 / ( \Delta f_\pi^4 r^6) $ and
attractive, necessarily being renormalizable with an arbitrary number
of counter-terms through energy dependent boundary
conditions~\cite{PavonValderrama:2007nu}. Clearly, the renormalization
of a potential where the $N\Delta$ splitting is treated as a small
scale deserves further investigation since it appears as an obvious
candidate where all necessary requirements for a convergent and short
distance insensitive scheme might be met. In the present paper we
check that the naive expectations based on the simple
Eq.~(\ref{eq:oppenheimer}) are indeed verified for the long distance
chiral potentials including $\Delta$ degrees of
freedom~\cite{Ordonez:1995rz,Kaiser:1997mw,Kaiser:1998wa,Krebs:2007rh}.

The delta-isobar has played a crucial role in the development of
nuclear and particle physics~(see e.g.~\cite{Ericson:1988gk,Cattapan:2002rx} 
and references therein).
Besides fixing the number of colours $N_c=3$ in QCD to comply with Pauli 
principle both at the quark as well as at the hadron level, 
this state can be clearly seen in $\pi N$ scattering as a resonance, 
and it ubiquitously appears whenever the nucleon excitation energy is about 
the Delta-Nucleon splitting, $\Delta= M_\Delta-M_N = 293 {\rm MeV}$. 
Already in the earliest implementations of Weinberg's
ideas the influence of Delta degrees of freedom was
considered~\cite{Ordonez:1992xp,VanKolck:1993ee,Ordonez:1995rz} in
old-fashioned perturbation theory where energy dependent potentials
are generated. Energy independent potentials have been obtained using
the Feynman diagram technique a decade
ago~\cite{Kaiser:1997mw,Kaiser:1998wa} and only recently the N2LO
contributions have also been worked out in Ref.~\cite{Krebs:2007rh},
where peripheral np phase shifts are computed in perturbation theory
for these $\Delta$ contributions.
Finite cut-off calculations involving $\Delta$-degrees of freedom to NLO have
been analyzed in momentum space~\cite{Epelbaum:1999dj}.
It has been shown that the inclusion of the $\Delta$-excitation improves
the convergence of the chiral expansion of the NN 
interaction~\cite{Pandharipande:2005sx,Epelbaum:2007sq}.
The renormalization of the $^1S_0$ phase with the NLO-$\Delta$ chiral
potential has also been discussed recently in momentum
space~\cite{Entem:2007jg}. Short distance insensitivity has also been
discussed in other contexts different from NN collisions such as $\pi d$ 
scattering at threshold~\cite{Valderrama:2006np} where a rationale
for the multiple scattering expansion was indeed supported by the
chiral singular potentials. Unfortunately, boost corrections being
proportional to the average relative pn kinetic energy in the
deuteron, $\langle p^2 \rangle_d /M $, turned out to diverge due to
the infinitely many short oscillations of the wave function.
Recently, the benefits of including the $\Delta$ in such a process are
discussed in \cite{Baru:2007wf} where it is shown how a proper
reorganization of the boost corrections not only makes them finite but
also numerically small after renormalization in harmony with
phenomenological expectations. The drawback is a proliferation of
low energy constants when including the $\Delta$, which can render 
the $\Delta$-full theory impractical at higher orders.
The implications of a $\Delta$-based power counting for the three nucleon 
problem are analyzed in Refs.~\cite{Pandharipande:2005sx,Epelbaum:2007sq}. 

In the present paper we focus our interest in the crucial role played
by the Delta-isobar in NN scattering in the elastic region from the
point of view of renormalization of chiral nuclear forces, and how it
might solve a long-standing problem. 
Besides an acceptable phenomenology in the s-wave phases we also show
that precisely because of the built-in short distance insensitivity and 
unlike previous calculations where the $\Delta$-degrees of freedom were 
absent, the deuteron is always bound up to N2LO-$\Delta$, 
the highest order computed at present~\cite{Krebs:2007rh}.

The paper is organized as follows. In Section~\ref{sec:form} we review
the formalism as applied to the energy independent potentials in a
$\Delta$-full theory~\cite{Kaiser:1997mw,Kaiser:1998wa,Krebs:2007rh}.
In Section~\ref{sec:singlet} we discuss the simpler $^1S_0$ channel 
including either one or two counter-terms by means of an energy dependent
boundary condition. We analyze the deuteron bound state and its
properties in Section~\ref{sec:triplet}. Scattering states in the
${}^3S_1-{}^3D_1$ channel as well as the corresponding phase shifts are
constructed by orthogonality to the deuteron and compared to the
results obtained with the Nijmegen II potential~\cite{Stoks:1994wp}, 
which has a $\chi^2$ per datum near one and therefore can
be considered as an alternative partial wave analysis, 
compatible with the original Nijmegen PWA~\cite{Stoks:1993tb}.
The results found in this paper are
further compared to the $\Delta$-less theory in
Section~\ref{sec:comparison} with previous
results~\cite{Valderrama:2005wv}. In Sec.~\ref{sec:spectral} we
address the issue of the relevant scales entering the calculation as
well as the role of missing effects such as relativistic, $3\pi$
contributions or vector meson exchange calculations.  We also
re-analyze a regularization of the potential based on the spectral
representation motivated by dispersion
relations~\cite{Epelbaum:2003gr,Epelbaum:2003xx,Epelbaum:2004fk,Krebs:2007rh}
and discuss its meaning on the light of the present approach. Finally
in Section~\ref{sec:conclusions} we recapitulate our results and draw
our main conclusions.

\section{Formalism}
\label{sec:form} 

In the present paper we use coordinate space renormalization by means
of boundary
conditions~\cite{PavonValderrama:2005gu,Valderrama:2005wv,PavonValderrama:2005uj}.
The equivalence to momentum space renormalization using counter-terms
for singular potentials is discussed thoroughly in
Refs.~\cite{Entem:2007jg,Valderrama:2007ja}.  We use the energy
independent long distance chiral potentials of
Ref.~\cite{Kaiser:1997mw,Kaiser:1998wa,Krebs:2007rh}, which are
on-shell equivalent to the energy-dependent potentials of
Ref.~\cite{Ordonez:1995rz}.  We follow the convention of taking $1/M$
corrections as ${\cal O}(p^2)$, in agreement with the original
argument from Ref.~\cite{Weinberg:1991um} as well as with the
calculations of Refs.~\cite{Ordonez:1995rz,Epelbaum:2004fk}, and the
computation of the N2LO-$\Delta$ potential of
Ref.~\cite{Krebs:2007rh}.  We remind that this convention is not
universal, and keeping the $1/M$ corrections as ${\cal O}(p)$ is also
customary~\footnote{Although the effect of keeping or ignoring the
  mass corrections is indeed small (see end of this section).}. This
convention has the consequence that at NLO-$\Delta$ and N2LO-$\Delta$
relativistic corrections are suppressed~\cite{Krebs:2007rh}.  The
final expressions for the potential are taken from
Ref.~\cite{Krebs:2007rh}.

\begin{table*}
%\begin{ruledtabular}
\begin{tabular}{|c|c|c|c|c|c|c|}
\hline 
 & $ M C_{6,V,C}^{\rm NLO-\Delta} $ & $ M C_{6,W,C}^{\rm NLO-\Delta} $ & $ M C_{6,V,S}^{\rm NLO-\Delta} $ & $ M C_{6,W,S}^{\rm NLO-\Delta} $ & $ M C_{6,V,T}^{\rm NLO-\Delta} $ & $ M C_{6,W,T}^{\rm NLO-\Delta} $ \\
\hline
Fit-1 & -7.233 & -1.201 & 0.601  & 0.301 & -0.601 & -0.301 \\ 
Fit-2 & -3.867 & -0.833 & 0.417  & 0.177 & -0.417 & -0.177 \\ 
\hline
\hline
 & $ M C_{6,V,C}^{\rm N2LO-\Delta} $ & $ M C_{6,W,C}^{\rm N2LO-\Delta} $ & $ M C_{6,V,S}^{\rm N2LO-\Delta} $ & $ M C_{6,W,S}^{\rm N2LO-\Delta} $ & $ M C_{6,V,T}^{\rm N2LO-\Delta} $ & $ M C_{6,W,T}^{\rm N2LO-\Delta} $ \\
\hline
Fit-1 & -1.241  & -0.489 & 0.244  & 0.351 & -0.244 & -0.351 \\ 
Fit-2 & -4.381  & -1.121 & 0.561  & 0.439 & -0.561 & -0.439  \\ 
\hline
\end{tabular}
%\end{ruledtabular}
\caption{\label{tab:c_6-VdW} van der Waals $ M C_6$ coefficients (in
${\rm fm}^4$) for the different spin-isospin components of the
NLO-$\Delta$ and N2LO-$\Delta$ potentials. We use the $\pi N$
motivated Fits 1 and 2 of Ref.~\cite{Krebs:2007rh}. Fit 1 involves the
SU(4) quark-model relation $h_A= 3 g_A /(2 \sqrt{2}) = 1.34 $ for
$g_A=1.26$.}
\end{table*}

We solve the coupled channel (coupled in angular momentum) 
Schr\"odinger equation for the relative motion, 
which in compact notation reads,
\begin{eqnarray}
\left[ - \frac{\nabla^2}{M } + 
{V}_{NN} (\vec x) \right] \,\Psi (\vec x) = E_{{\rm c.m.}}\,\Psi (\vec x) \, ,
\label{eq:sch_cp} 
\end{eqnarray} 
where the spin and isospin indices have not been explicitly written;
$M = 2 M_p M_n /(M_p+M_n)$ is twice the reduced proton-neutron mass. 
In coordinate space the potential can be written as
\begin{eqnarray}
{V}_{NN} (\vec x)&=& V_C (r) + \tau \,W_C (r) \nonumber\\ &+& 
\sigma \left( V_{S} (r) + \tau \,W_{S} (r) \right) \nonumber \\ &+& 
 S_{12} \left( V_T (r) + \tau \,W_T (r) \right) \, ,
\end{eqnarray}
where spin-orbit and quadratic spin-orbit terms have been ignored as they
are not present in the NLO-$\Delta$ and N2LO-$\Delta$ 
potentials from Ref.~\cite{Krebs:2007rh}. 
The operators $\tau$, $\sigma$ and $S_{12}$ are given by
\begin{eqnarray}
\tau &=& \vec \tau_1 \cdot
\vec \tau_2 = 2 t(t+1) -3 \, , \nonumber \\ 
\sigma &=& \vec \sigma_1 \cdot \vec \sigma_2 = 2 s(s+1) -3 \, ,
\nonumber \\ 
S_{12} &=& 3\,\vec \sigma_1 \cdot \hat r \vec \sigma_2 \cdot \hat
r - \vec\sigma_1 \cdot \vec \sigma_2 \, ,
\end{eqnarray} 
where $\vec{\tau}_{1(2)}$ and $\vec{\sigma}_{1(2)}$ represent the
proton(neutron) isospin and spin operators; $\vec \tau_1 \cdot \vec
\tau_2$ and $\vec \sigma_1 \cdot \vec \sigma_2$ are evaluated for
total isospin $t=0,1$ and total spin $s=0,1$.  Note that in
Ref~\cite{Krebs:2007rh} the potential is local and there are no
relativistic mass corrections at N2LO-$\Delta$.  For states with good
total angular momentum $j$ and total spin $s = 1$, the tensor operator
reads
\begin{eqnarray}
S_{12}^j &=& \begin{pmatrix} 
-\frac{2(j-1)}{2j+1} & 0 & \frac{6\sqrt{j(j+1)}}{2j+1} \\ 
0 & 2 & 0 \\
\frac{6\sqrt{j(j+1)}}{2j+1} & 0 & -\frac{2(j+2)}{2j+1} 
\end{pmatrix} \, ,
\end{eqnarray} 
where the matrix indices represent the orbital angular momentum 
$l=j-1, j, j + 1$. 
For $s = 0$ states the tensor operator vanishes $S_{12}^j = 0$.
We remind that Fermi-Dirac statistics implies $(-)^{l+s+t}=-1$.

A remarkable feature which happens at LO, NLO-$\Delta$ and
N2LO-$\Delta$~\cite{Ordonez:1995rz,Kaiser:1997mw,Kaiser:1998wa,Krebs:2007rh}
is that the spin-orbit coupling vanishes,
as well as any relativistic corrections. 
As a consequence, for a given isospin the potential can be diagonalized.
The corresponding eigen-potentials depend just on the isospin of the channel 
and not on the total angular momentum $j$. All the $j$ dependence goes
into the matrix which diagonalizes the potential. This can be seen in
the following formulas:
\begin{eqnarray}
{\bf V}^{1j} &=& (V_C + \tau W_C) - 3\,(V_S + W_S \tau ) + 
S_{12}^j\,(V_T +\tau W_T) \nonumber \\ &=& 
M_j\,{\rm diag} \left( A -4 B, A + 2 B, A + 2 B \right)\,M_j^{-1} \, ,
\end{eqnarray} 
where ${\bf V}^{1j}$ is the triplet-channel potential
written in matrix form, with 
\begin{eqnarray}
A &=& (V_C + \tau W_C) - 3\,(V_S + W_S \tau )\, , \nonumber \\
B &=& (V_T +\tau W_T)\, ,  \nonumber \\
M_j &=& \begin{pmatrix}   
\cos \theta_j & 0 & -\sin \theta_j \\ 
0 & 1 & 0 \\
\sin \theta_j & 0 &  \cos \theta_j 
\end{pmatrix} \, ,
\end{eqnarray} 
where 
\begin{eqnarray}
\cos \theta_j =  \sqrt{\frac{j}{2j+1}}
\label{eq:cos-theta} \, .
\end{eqnarray} 
The transformation $M_j$ diagonalizes the full potential but {\it not}
the Schr\"odinger equation, since it contains in addition to the potential
the centrifugal barrier, which is a diagonal operator in the $jls$ basis,
but does not remain diagonal in the rotated basis. 
Specific knowledge of the short distance behaviour is needed to carry on 
with the renormalization program. 
Generally, on purely dimensional grounds, we have for $r\to 0$
\begin{eqnarray}
V_i (r) \to \frac{C_{k}^{V,i}}{r^k} \, , \qquad W_i (r) \to
\frac{C_{k}^{W,i}}{r^k} \, ,
\end{eqnarray} 
where
\begin{eqnarray}
C_{k = 2n+m+r+1}^i \sim \frac1{ f_\pi^{2n} M_N^m \Delta^r} \, ,
\end{eqnarray} 
with $\Delta$ the $N\Delta$ splitting, and $n$, $m$ and $r$
nonnegative integers.  At short distances, the angular momentum
dependence may be neglected when the index $k > 2$.  The relevant
issue to carry out the renormalization procedure and to generate
finite results is to know whether the interaction is attractive or
repulsive.  It turns out that both NLO-$\Delta$ and N2LO-$\Delta$
potentials have a leading singularity behaviour of $1/r^6$.  In
Appendix~\ref{appendix} we list the analytical expressions for the van
der Waals coefficients $C_k^i$, for $k=6$ (i.e. the leading
singularity of the potential).

In our numerical calculations we take $f_\pi=92.4 {\rm MeV}$,
$m_\pi=138.03 {\rm MeV}$, $ 2 \mu_{np}= M_N = 2M_p M_n /(M_p+M_n) =
938.918 {\rm MeV}$, $g_A =1.29$ in the OPE piece to account for the
Goldberger-Treimann discrepancy and $g_A=1.26$ in the TPE piece of 
the potential~\footnote{Strictly speaking $g_A = 1.26$
both in the OPE and TPE pieces
of the potential, but it happens that at NLO and higher orders the OPE piece
receives a contribution from the $d_{18}$ LEC, related to the 
Goldberger-Treimann discrepancy, see, for example, the expressions of
Ref.~\cite{Epelbaum:2003gr} for details.
This is equivalent to consider the original expression for the OPE potential,
but taking $g_A = 1.29$ instead of $g_A = 1.26$.
}.
The corresponding pion nucleon coupling constant takes 
then the value $ g_{\pi NN}=13.083$ for the OPE piece of the potential.
We use the $\Delta N$ splitting $\Delta = 293\,{\rm MeV}$.
For $h_A$ and the low energy constants $c_1$, $c_2$, $c_3$, $c_4$, 
$b_3$ and $b_8$ we take the values from Fit 1 and 2 of 
Ref.~\cite{Krebs:2007rh} (table I within that reference),
where they are deduced from a fit to $\pi N$ threshold parameters in S- and
P-waves to the data of Ref.~\cite{Matsinos:1997pb}.
These values are compatible with all $\pi N$ threshold parameters 
except $b_{0,+}^-$ which is about twice its recommended 
value~\cite{Matsinos:1997pb} at this level of approximation. 
Fit 1 uses the SU(4) quark-model relation $h_A = 3\,g_A / 2\sqrt{2}$
($=1.34$ for $g_A = 1.26$), while Fit 2 uses $h_A = 1.05$.

The numerical values of the van der Waals coefficients $M C_6$
are summarized in Table~\ref{tab:c_6-VdW} for the different 
components of the NLO-$\Delta$ and N2LO-$\Delta$ potentials.
The van der Waals coefficients are additive: therefore one can
obtain the coefficient corresponding to some given partial wave
by adding the individual contributions, i.e.
\begin{eqnarray}
M C_6 &=& M C_{6,V,C} (r) + \tau \,M C_{6,W,C} \nonumber\\ &+& 
\sigma \left( M C_{6,V,S} + \tau \,M C_{6,W,S} \right)\nonumber \\ &+& 
S_{12} \left( M C_{6,V,T} + \tau \,M C_{6,W,T} \right) \, .
\end{eqnarray}
This is done for the singlet $^1S_0$ and the triplet $^3S_1-{}^3D_1$ channels
in Table~\ref{tab:c_6-channels}. 

We also present N2LO-$\Delta \negate$ results for comparison purposes.
For them we use the same parameters as in the N2LO-$\Delta$ computation,
except for the $c_1$, $c_3$ and $c_4$ LECs, for which we use the following
two different determinations: the so-called set IV of 
Ref.~\cite{Valderrama:2005wv} 
%($c_1 = -0.81\,{\rm GeV}^{-1} , 
%\,c_3 = -3.20\,{\rm GeV}^{-1} , \, c_4 = 5.40\,{\rm GeV}^{-1}$)
, which 
was obtained in Ref.~\cite{Entem:2003ft} by fitting to NN scattering data, 
and the values from Ref.~\cite{Krebs:2007rh} for N2LO-$\Delta \negate$
%($c_1 = -0.57\,{\rm GeV}^{-1} , \,c_3 = -3.87\,{\rm GeV}^{-1} , 
%\, c_4 = 2.89\,{\rm GeV}^{-1}$), 
which we refer to as set $\pi$N 
and allow a better comparison
with the N2LO-$\Delta$ computations presented here, as they are also
fitted to reproduce $\pi N$ S- and P-wave threshold parameters.
Due to the different fitting procedures, any direct comparison between 
the results of Ref.~\cite{Valderrama:2005wv}, i.e. set IV, and 
the N2LO-$\Delta$ results should be done with care.
It should be mentioned too that the results of Ref.~\cite{Valderrama:2005wv}
contain 1/M corrections to the potential.
These corrections are nevertheless small and, if excluded, would only induce 
small differences in the results of Ref.~\cite{Valderrama:2005wv}~\footnote{
As an illustration, ignoring the 1/M corrections will yield to the following
modifications for the results of Ref.~\cite{Valderrama:2005wv} (original
results in parentheses): $A_S = 0.887(0.884) \, {\rm fm}^{1/2}$, 
$r_m = 1.972(1.967)\,{\rm fm}$, 
$Q_d = 0.278(0.276)\,{\rm fm}^2$, 
$P_D = 8(8)\,\%$, 
$\langle r^{-1} \rangle = 0.442(0.447)\,{\rm fm}^{-1}$, 
$\langle r^{-2} \rangle = 0.276(0.284)\,{\rm fm}^{-2}$. 
}.

\begin{table*}
%\begin{ruledtabular}
\begin{tabular}{|c|c|c|c|c|c|c|}
\hline 
 & $ M C_{6,^1S_0}^{\rm NLO-\Delta} $ & $ M C_{6,^3S_1}^{\rm NLO-\Delta} $ & $ M C_{6,E_1}^{\rm NLO-\Delta} $ & $ M C_{6,^3D_1}^{\rm NLO-\Delta} $ & $ -R_+^4 $ & $ -R_-^4 $ \\
\hline
Fit 1 & -11.138  & -3.932 & 0.855  & -4.537 & -5.141 & -3.327 \\ 
Fit 2 & -6.481  & -1.482 & 0.322  & -1.710 & -1.938 & -1.254 \\ 
\hline
\hline
 & $ M C_{6,^1S_0}^{\rm N2LO-\Delta} $ & $ M C_{6,^3S_1}^{\rm N2LO-\Delta} $ & $ M C_{6,E_1}^{\rm N2LO-\Delta} $ & $ M C_{6,^3D_1}^{\rm N2LO-\Delta} $ & $ -R_+^4 $ & $ -R_-^4 $ \\
\hline
Fit-1 & -3.519   & -0.585 & 2.290  & -2.204 & -3.823 & 1.035 \\ 
Fit-2 & -8.500 & -1.773 & 2.138  & -3.284 & -4.796 & -0.261 \\ 
\hline
\end{tabular}
%\end{ruledtabular}
\caption{\label{tab:c_6-channels} van der Waals $ M C_6$ coefficients
(in ${\rm fm}^4$) in the $^1S_0$ and $^3S_1-{}^3D_1$ (deuteron) channels
of the NLO-$\Delta$ and N2LO-$\Delta$ potentials. We also present the
corresponding negative eigenvalues $-R_+^4$ and $-R_-^4$ in the
deuteron channel case. We use the $\pi N$ motivated Fits 1 and 2 of
Ref.~\cite{Krebs:2007rh}. Fit 1 involves the SU(4) quark-model
relation $h_A= 3 g_A /(2 \sqrt{2}) = 1.34 $ for $g_A=1.26$.}
\end{table*}

\section{The singlet channel}
\label{sec:singlet}

\subsection{Equations and boundary conditions}

The $^1S_0$ wave function in the pn center-of-mass (c.m.) system
can be written as
\begin{eqnarray}
\Psi (\vec x) &=& \frac1{\sqrt{4\pi}\, r } \, u(r) \chi_{pn}^{s m_s} \, ,
\end{eqnarray} 
with the total spin $s=0$ and $m_s=0 $.  The function $u(r)$ is the
reduced S- wave function, and satisfies the following reduced 
Schr\"odinger equation
\begin{eqnarray}
-u_k '' (r) + U_{^1S_0} (r) u_k (r)  &=& k^2 u_k (r) \, ,
\label{eq:sch_singlet} 
\end{eqnarray}
where $k$ is the center-of-mass momentum, and $U_{^1S_0}$ the reduced potential
defined as
\begin{eqnarray}
U_{^1S_0}(r) &=& M\,\Big( V_C(r) + W_C(r) \nonumber\\
&& - 3 V_S(r) - 3 W_S(r)\Big) \, .   
\end{eqnarray} 
For a finite energy scattering state we solve for the chiral
potential with the asymptotic normalization
\begin{eqnarray}
u_k (r) \to \frac{\sin(k r + \delta_0(k))}{\sin \delta_0(k)} \, , 
\label{eq:norm}
\end{eqnarray} 
with $\delta_0(k)$ the phase shift. For a potential falling off
exponentially $\sim e^{-m_\pi r}$ at large distances, we have
the effective range expansion, valid for momenta $|k| < m_\pi/2$,
\begin{eqnarray}
k \cot \delta_0 (k) = - \frac1\alpha_0 + \frac12 r_0 k^2 + v_2 k^4 + \dots 
\end{eqnarray} 
with $\alpha_0$ the scattering length and $r_0$ the effective range.
At short distances the NN chiral potential behaves as
\begin{eqnarray} 
U_{^1 S_0} (r) & \to &  \frac{M C_{6,^1S_0}}{r^6} = - \frac{R^4}{r^6} \, ,
\label{eq:pot_sing_short}
\end{eqnarray} 
where 
\begin{eqnarray}
M C_{6,^1S_0} &=& M C_{6,^1S_0}^{\rm NLO-\Delta}  + 
M C_{6,^1S_0}^{\rm N2LO-\Delta} \, , 
\end{eqnarray} 
which is a van der Waals type interaction ~\footnote{It should be
  noted that the NLO-$\Delta \negate$ potential is less singular than
  the NLO-$\Delta$ one at short distances, diverging as $1/r^5$
  only.}.  The numerical values for $M C_{6,^1S_0}^{\rm NLO-\Delta}$
and $M C_{6,^1S_0}^{\rm N2LO-\Delta}$ are listed in
Tab.~\ref{tab:c_6-channels}.  The value of the coefficient is
negative, with the typical length scale $R=(-M C_6)^{1/4}$.  The
solution at short distances is of oscillatory type:
\begin{eqnarray}
u_k (r) &\to & A \left(\frac{r}{R}\right)^{3/2} 
\sin \left[ \frac{1}{2} \left(\frac{R}{r}\right)^{2} + \varphi  \right] \, ,
\end{eqnarray} 
where $A$ is a normalization constant and $\varphi$ an undetermined phase, 
which may in principle depend on energy. 

We will present below two calculations. In the first one the
renormalization is carried out with one counter-term, which in turn
means one renormalization condition. In such a case the short distance
phase becomes energy independent and orthogonality conditions are
satisfied. In the second calculation we proceed with two counter-terms,
i.e. two renormalization conditions, for which the short distance
phase acquires a very specific energy dependence.

\begin{figure*}[]
\begin{center}
\epsfig{figure=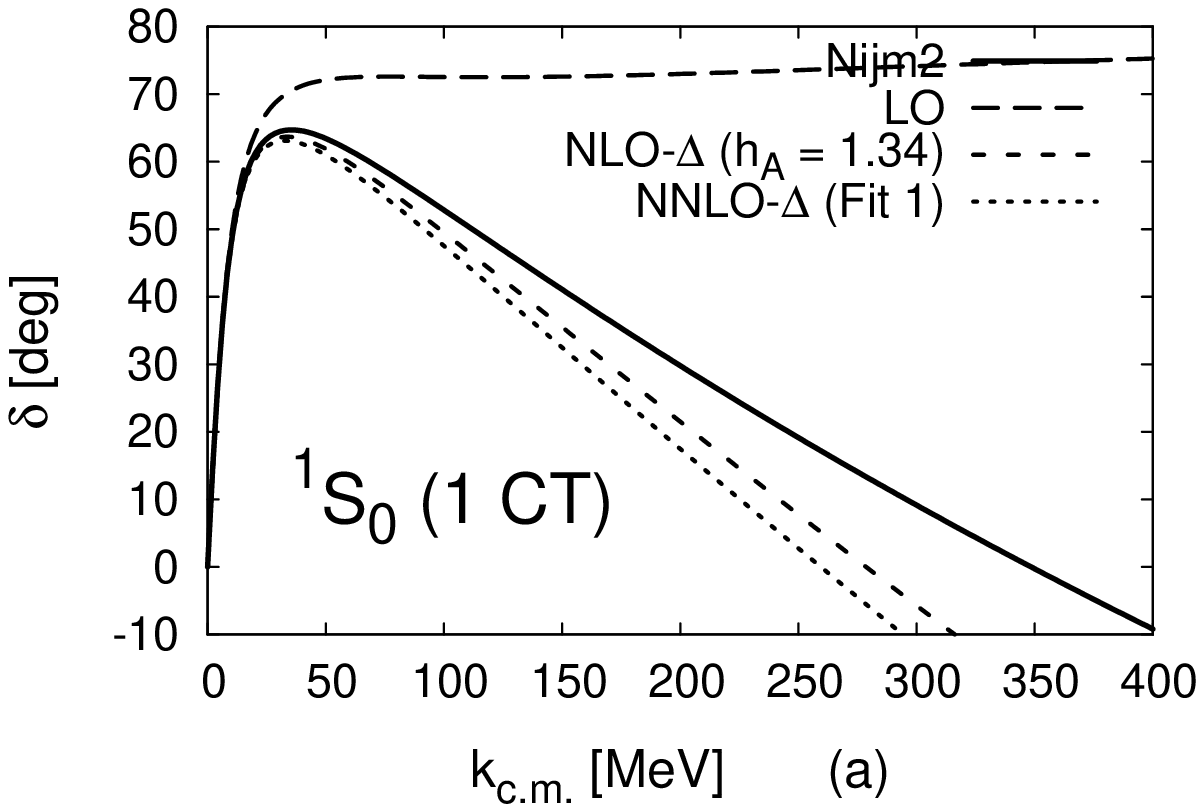,height=6.5cm,width=6.5cm}\hskip1.5cm
\epsfig{figure=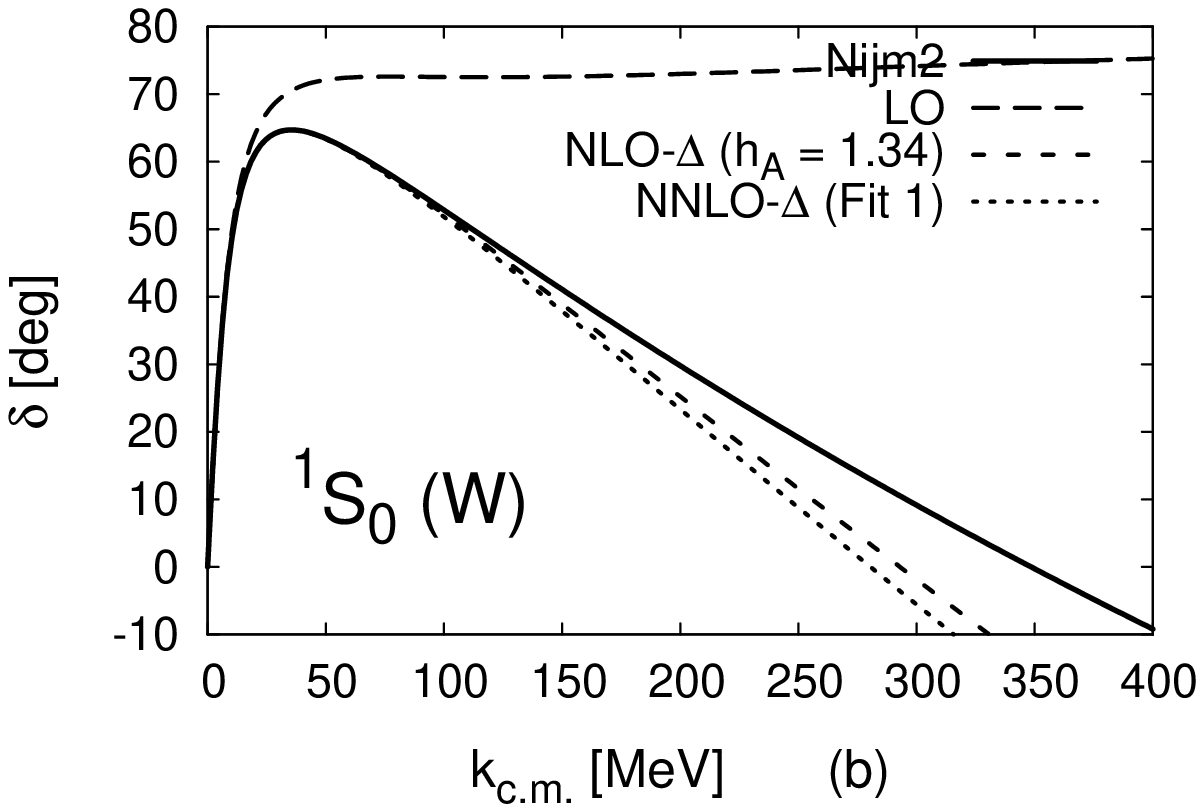,height=6.5cm,width=6.5cm}
\epsfig{figure=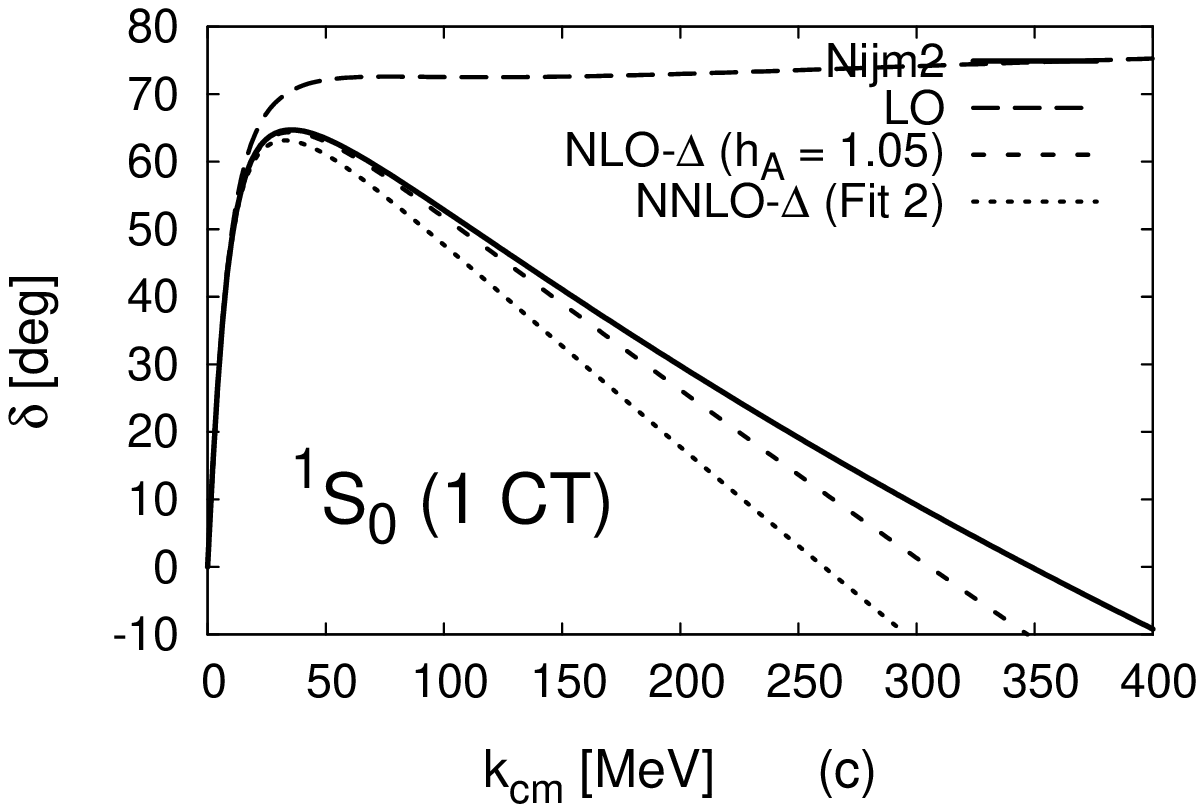,height=6.5cm,width=6.5cm}\hskip1.5cm
\epsfig{figure=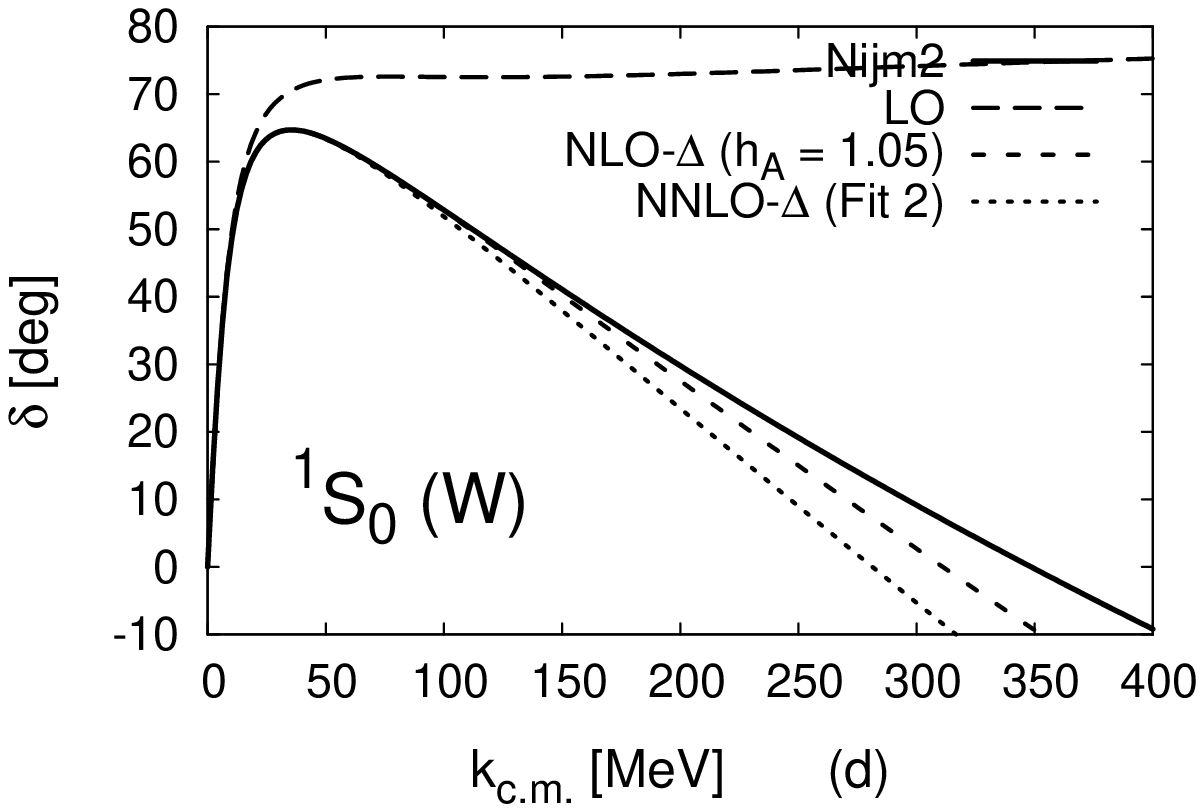,height=6.5cm,width=6.5cm}
\end{center}
\caption{Renormalized phase shifts for the LO, NLO-$\Delta$ and
N2LO-$\Delta$ potentials as a function of the c.m. np momentum 
$k_{\rm c.m.}$ in the $^1S_0$ singlet channel,
compared to the Nijmegen II potential results~\cite{Stoks:1994wp}.
The computations are done both with $h_A = 1.34$ (upper panels) and
with $h_A = 1.05$ (lower panels).
For the LO result, the calculation is always done with one counterterm,
while for the NLO and N2LO results, computations are done with one
counter-term (left panels) and two counter-terms (right panels),
corresponding this last case to the standard Weinberg counting
for the $^1S_0$ channel.}
\label{fig:phase-shift_1S0}
\end{figure*}

\subsection{Renormalization with one counter-term} 

As mentioned, the phase shift is determined from Eq.~(\ref{eq:norm}),
but to fix the undetermined phase $\varphi$ we impose orthogonality
for $r > r_c$ between the zero energy state and the state with
momentum $p$. As shown in~\cite{Valderrama:2005wv}, orthogonality
turns out to be equivalent to the following condition between 0- and 
k-momentum reduced wave functions at $r = r_c$
\begin{eqnarray} 
u_k'(r_c)  u_0 (r_c) - u_0'(r_c) u_k(r_c) =0 \, . 
\label{eq:orth_singlet}
\end{eqnarray} 
Taking the limit $r_c \to 0$ implies that the short distance phase
$\varphi$ is energy independent~\cite{Valderrama:2005wv}. Thus, for
the zero energy state we solve
\begin{eqnarray}
-u_0 '' (r) + U_{^1S_0} (r) u_0 (r)  &=& 0 \, ,
\label{eq:sch_singlet_0} 
\end{eqnarray}
with the asymptotic normalization at large distances
\begin{eqnarray} 
u_0 (r) &\to & 1- \frac{r}{\alpha_0}  \, ,
\end{eqnarray}
where $\alpha_0$ is the scattering length. In this equation $\alpha_0$
is an input, so one needs to integrate Eq.~(\ref{eq:sch_singlet_0}) from
infinity to the origin (in contrast with the usual procedure of 
integrating from the origin to infinity). 
The effective range, defined as
\begin{eqnarray} 
r_0 &=& 2 \int_0^\infty dr \left[ \left(1-\frac{r}{\alpha_0} \right)^2-
u_0 (r)^2 \right] 
\label{eq:r0_singlet} \, ,
\end{eqnarray} 
can be computed. 
Due to the superposition principle, we can write the zero momentum wave
function as the following linear combination
\begin{eqnarray} 
u_0 (r) &=& u_{0,c} (r) - \frac1{\alpha_0} u_{0,s} (r) \, ,  
\end{eqnarray} 
where $u_{0,c} (r) \to 1 $ and $ u_{0,s} (r) \to r $ correspond
to cases where the scattering length is either infinity or zero
respectively. Using this decomposition, we get
\begin{eqnarray} 
r_0  &=&  A + \frac{B}{\alpha_0}+ \frac{C}{\alpha_0^2}  \, ,    
\label{eq:r0_univ} 
\end{eqnarray} 
where $A$, $B$ and $C$, defined as:
\begin{eqnarray}
A &=& 2 \int_0^\infty dr ( 1 - u_{0,c}^2 ) \, , \\    
B &=& -4 \int_0^\infty dr ( r - u_{0,c} u_{0,s} ) \, , \\    
C &=& 2 \int_0^\infty dr ( r^2 - u_{0,s}^2 )    \, ,  
\end{eqnarray} 
depend on the potential parameters only. The interesting thing is that
all dependence on the scattering length $\alpha_0$ is displayed
explicitly by Eq.~(\ref{eq:r0_univ}).  Numerically we get 
\begin{eqnarray}
r_0&=& 2.661 - \frac{5.707}{\alpha_0} + \frac{5.988}{\alpha_0^2}
\quad ({\rm NLO-}\Delta, {\rm h_A = 1.34})\, , \nonumber \\
r_0&=& 2.517 - \frac{5.430}{\alpha_0} + \frac{5.811}{\alpha_0^2}
\quad ({\rm NLO-}\Delta, {\rm h_A = 1.05})\, , \nonumber \\  
r_0 &=& 2.789 - \frac{5.992}{\alpha_0} + \frac{6.187}{\alpha_0^2}
\quad ({\rm N2LO-}\Delta, {\rm Fit 1})\, ,\nonumber \\  
r_0 &=& 2.780 - \frac{5.969}{\alpha_0} + \frac{6.171}{\alpha_0^2}
\quad ({\rm N2LO-}\Delta, {\rm Fit 2})\, .\nonumber \\  
\end{eqnarray} 
The corresponding numerical values when the experimental $\alpha_0 = -
23.74 {\rm fm} $ is taken, as well as the $v_2$ parameter can be looked
up in Table~\ref{tab:table_singlet}. As can be seen, the value of the
effective range has a clear convergence pattern when going from LO to 
N2LO-$\Delta$. The contribution coming from N2LO-$\Delta$ is very 
small compared to the NLO-$\Delta$ contribution, in agreement with
the findings of Ref.~\cite{Krebs:2007rh}. This is in contrast with
the $\Delta$-less theory, for which N2LO generates a great correction
over the NLO results. 
Unfortunately the N2LO-$\Delta$ results converge to wrong values, 
about $3\,{\rm fm}$ for both Fit 1 and 2. The same happens for the
N2LO-$\Delta \negate$ results, although with a weaker convergence
pattern.
The reason for this discrepancy is that the NLO and N2LO potentials,
both in $\Delta$-less and $\Delta$-full theories, are too attractive 
at intermediate distances, thus yielding a bigger value for $r_0$ 
than the one obtained with phenomenological potentials like 
Nijmegen II or Reid93~\cite{Valderrama:2005wv,Entem:2007jg}.

\begin{table}[ttt] 
\caption{\label{tab:table_singlet} Predicted threshold parameters in
the singlet $^1S_0 $ channel with one counter-term for Fits 1 and 2 
of Ref.~\cite{Krebs:2007rh}. 
We compare our renormalized results given
by the cut-off independent universal formula (\ref{eq:r0_univ}) for
$r_0$ and its extension for $v_2$ to finite cut-off NN calculations
using their scattering length as an input. The experimental values
for the scattering length and effective range are taken from 
Ref.~\cite{Machleidt:2000ge}.}
\begin{ruledtabular}
\begin{tabular}{|c|c|c|c|c|}
\hline  & Calculation & $\alpha_0 ({\rm
fm}) $ & $r_0 ({\rm fm}) $ & $v_2 ({\rm fm}^3 ) $  \\ \hline
LO  &  Ref.~\cite{Valderrama:2005wv}  &  Input  &   1.44  & -2.11  \\ \hline 
NLO-$\Delta \negate$  &  Ref.~\cite{Valderrama:2005wv} &  Input  &   2.29  & -1.02  \\ \hline
NLO-$\Delta$ ($h_A = 1.34$)&  This work &  Input  &   2.91  & -0.32 \\
NLO-$\Delta$ ($h_A = 1.05$) &  This work &  Input  &   2.76  & -0.53 \\  \hline 
%N2LO-$\Delta \negate$ Set I  & Ref.  & Input  & 2.92 & -0.30 \\ 
%N2LO-$\Delta \negate$ Set II & Ref.  & Input & 2.97 & -0.23 \\   
%N2LO-$\Delta \negate$ Set III & Ref. & Input  & 2.83 & -0.43 \\    
N2LO-$\Delta \negate$ (Set IV) & Ref.~\cite{Valderrama:2005wv} & Input   & 2.87 & -0.38   \\ \hline
% N2LO-$\Delta \negate$ (Set IV) & This work & Input   & 2.94 & -0.29 \\
N2LO-$\Delta \negate$ ($\pi$N) & This work & Input   & 2.92 & -0.31 \\ \hline 
N2LO-$\Delta$ (Fit 1) &  This work &  Input  &   3.05  &  -0.12 \\ 
N2LO-$\Delta$ (Fit 2) &  This work &  Input  &   3.04  &  -0.13 \\ \hline
Nijm II  & Ref.~\cite{Stoks:1993tb,Stoks:1994wp}   & -23.73   & 2.67 & -0.48   
\\ 
Reid 93  & Ref.~\cite{Stoks:1993tb,Stoks:1994wp}   & -23.74   & 2.75 & -0.49   \\ \hline 
Exp. & --& -23.74(2) & 2.77(5) & - 
\end{tabular}
\end{ruledtabular}
\end{table}

%\subsection{Phase shift}

Using the orthogonality condition, Eq.~(\ref{eq:orth_singlet}), the
phase-shift can be determined from the scattering length and the
potential as independent parameters when the limit $r_c \to 0$ is
taken. The renormalized phase shift is presented in
Fig.~\ref{fig:phase-shift_1S0} (left).
The phase shifts are compared with the ones obtained from the Nijmegen II
potential~\cite{Stoks:1994wp}, which are compatible with the Nijmegen
PWA~\cite{Stoks:1993tb}.
As we see the trend in the
effective range $r_0$ and the $v_2$ parameter is reflected in the
behavior of the phase shift. In Ref.~\cite{Entem:2007jg} a similar
calculation was carried out in momentum space with inclusion of
$\Delta$ degrees of freedom at NLO and one counter-term. 
The present NLO-$\Delta$ coordinate space results agree with 
that calculation when the SU(4) relation, $h_A = 1.34$, is used.
As discussed in the previous paragraph for the effective range parameters,
at N2LO-$\Delta$ the phase shifts have already converged, although to the
wrong value, due to the excessive strength of the intermediate range of
the potential.
In the next section we will see how the situation changes when an
extra counter-term is included in the computations. This counter-term
will be fitted to reproduce the effective range.

\subsection{Renormalization with two counter-terms}

In the standard Weinberg counting both the long distance potential as
well as the short distance potential are dimensionally expanded in
momentum space. For the short range piece, one writes $V_S(p',p) = C_0
+ C_2 (p^2 + p'^2 ) + \dots\,$, where $C_0$, $C_2$, etc, are referred
to as counter-terms. 
In Ref.~\cite{Entem:2007jg} the
interrelation between the counter-terms and short distance boundary conditions 
in coordinate space has been discussed at length. 
At LO in the Weinberg counting one has only one
counter term. This corresponds to the situation described in the
previous section. At NLO and N2LO in the Weinberg counting
(both in the $\Delta$-less and $\Delta$-full cases), 
one adds two counter-terms $C_0$ and $C_2$ for a fixed
momentum space cut-off $\Lambda$, which may be fixed by reproducing
the scattering length $\alpha_0$ and the effective range $r_0$.
The interesting finding was that such a procedure {\it does not} generate
a converging $^1S_0$ phase shift in the limit $\Lambda \to
\infty$~\cite{Entem:2007jg}. 
Thus, the momentum space polynomial parameterization of Weinberg counting is 
not compatible with renormalization. However, this does not necessarily mean 
that one cannot impose two renormalization conditions to fix the scattering
length and the effective range.
Actually, on a wider perspective one may pose on the one hand the problem of 
obtaining a finite phase shift embodying the chiral potential and on the other
the problem of fixing both $\alpha_0$ and $r_0$ as 
{\it independent} parameters. 
Fortunately, there exists a unique procedure in coordinate 
space~\cite{PavonValderrama:2007nu} meeting the two previous conditions 
and  yielding convergent phase shifts, which we apply below to discuss 
the $^1S_0$  channel when two counter-terms are considered for NLO-$\Delta$ and 
N2LO-$\Delta$ chiral potentials.
We will refer to this scheme as Weinberg counting with $\Delta$.

The equivalent coordinate space procedure~\cite{PavonValderrama:2007nu,
Entem:2007jg} consists of expanding the wave function in powers of the
energy
\begin{eqnarray}
u_k (r) = u_0(r) + k^2 u_2 (r) + \dots
\end{eqnarray} 
where $u_0(r)$ and $u_2(r)$ satisfy the following equations, 
\begin{eqnarray} 
-u_0''(r) + U_{^1S_0} (r) u_0(r) &=& 0 \, , \label{eq:u0} \\  
u_0 (r) &\stackrel{r \to \infty}{\to}& 1 - \frac{r}{\alpha_0} \, ,  \nonumber \\ 
-u_2 '' (r) + U_{^1S_0} (r) u_2 (r) &=& u_0 (r) \, , \label{eq:u2} \\
u_2 (r) &\stackrel{r \to \infty}{\to}& \frac{\left(r^3 -3 \alpha_0\,r^2 + 3 \alpha_0 r_0\,r \right)}
{6 \alpha_0} \, . \nonumber
\end{eqnarray} 
The asymptotic conditions correspond to fixing $\alpha_0$ and $r_0$ as
independent parameters (two counter-terms). The matching condition at
the boundary $r=r_c$ becomes energy dependent~\cite{PavonValderrama:2007nu}
\begin{eqnarray}
\frac{u'_k (r_c)}{u_k(r_c)} = \frac{u'_0 (r_c) + k^2 u'_2 (r_c)+ \dots}{
u_0(r_c)+ k^2 u_2(r_c)+ \dots} \,\, ,
\label{eq:Lp}
\end{eqnarray} 
whence the corresponding phase shift may be deduced by integrating in
Eq.~(\ref{eq:u0}) and Eq.~(\ref{eq:u2}) and integrating out
Eq.~(\ref{eq:sch_singlet}) with Eq.~(\ref{eq:norm}). It is worth
mentioning that the energy dependent matching condition,
Eq.~(\ref{eq:Lp}), is quite unique since this is the only
representation for the boundary condition guaranteeing the existence 
of the limit $r_c \to 0 $
for singular potentials~\cite{PavonValderrama:2007nu}. As pointed out
in Refs.~\cite{PavonValderrama:2007nu, Entem:2007jg}, polynomial
expansions in $k^2$ such as suggested e.g. in the Nijmegen 
PWA~\cite{Stoks:1993tb}, and implemented later on for chiral TPE
potentials~\cite{Rentmeester:1999vw}, do not work for $r_c\to 0$, and
in fact generate undesired oscillations for $r_c \ll 1.4 {\rm fm}$ 
in the phase-shifts. 
The validity of these features can be deduced analytically in coordinate space 
as a consequence of the RG-invariance of a Moebius bilinear
transformation~\cite{PavonValderrama:2007nu}. 
Equivalent parameterizations in momentum space may likely exist, but are so 
far unknown. As already mentioned, the widely used polynomial
representation of short distance interactions in momentum space
$V_S(p',p) = C_0 + C_2 (p^2 + p'^2 ) + \dots $ of standard NLO and
NNLO Weinberg counting implies that for $\Lambda \to \infty$ either
$C_2 $ is irrelevant when only $\alpha_0$ is kept fixed or the phase
shift does not converge when both $\alpha_0$ and $r_0$ are
simultaneously fixed~\cite{PavonValderrama:2007nu, Entem:2007jg}.

In Fig.~\ref{fig:phase-shift_1S0} (right panel) we show the results
for the two counter-term renormalized phase shift at NLO-$\Delta$ and
N2LO-$\Delta$. As we see, the second counter-term is responsible for a
certain improvement all over the elastic region, in particular in the
region $k < m_\pi$ where TPE effects are expected to dominate in the
singlet $^1S_0$-channel~\footnote{In principle for $k < m_{\pi}$ OPE
  (instead of TPE) effects should dominate, but in the case of the
  $^1S_0$ singlet channel this is not the case due to the weakness of
  OPE in this channel. This only happens in the singlet channel; in
  the case of the triplet channel we recover the naive expectations,
  and OPE clearly dominates for $k < m_{\pi}$.}.

At higher momenta, however, the discrepancy with the Nijmegen 
II potential results~\cite{Stoks:1994wp} and therefore
with the Nijmegen PWA~\cite{Stoks:1993tb} persists.
The second counter-term is therefore unable to provide the
necessary repulsion to compensate for the excessive intermediate
range attraction present in the NLO-$\Delta$ and N2LO-$\Delta$ potentials. 
This result agrees with the findings of Ref.~\cite{Entem:2007jg} and 
extends them from NLO-$\Delta$ to N2LO-$\Delta$.
Possible solutions to this disturbing situation include finite cut-off 
computations, which help to increase the effect of the second counter-term, 
and the inclusion of spectral function regularization, 
proposed in Ref.~\cite{Epelbaum:2003gr}
precisely to treat the problem of the intermediate strength of the 
N2LO-$\Delta \negate$ potential in peripheral partial waves. 
The case of spectral function regularization will be discussed in 
Section \ref{sec:spectral}.

\begin{figure*}[ttt]
\begin{center}
\epsfig{figure=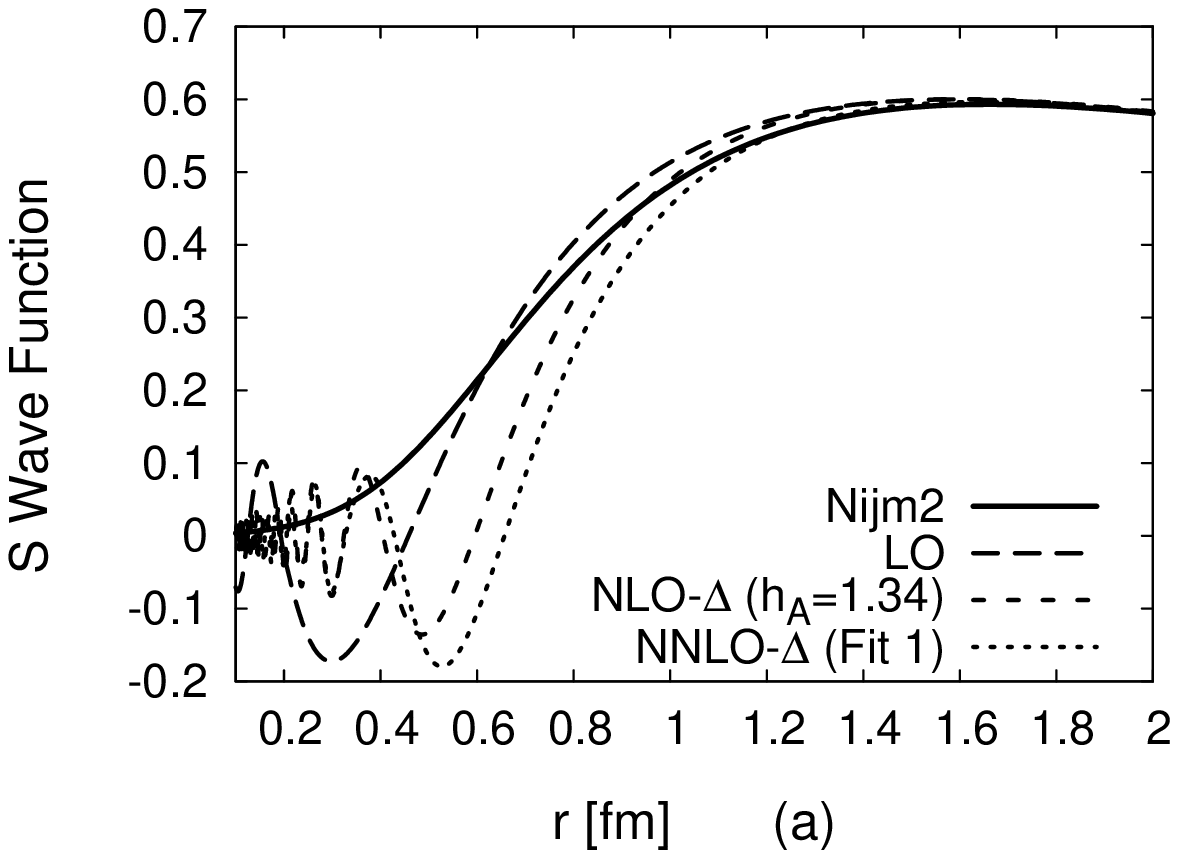,height=5cm,width=7cm}
\epsfig{figure=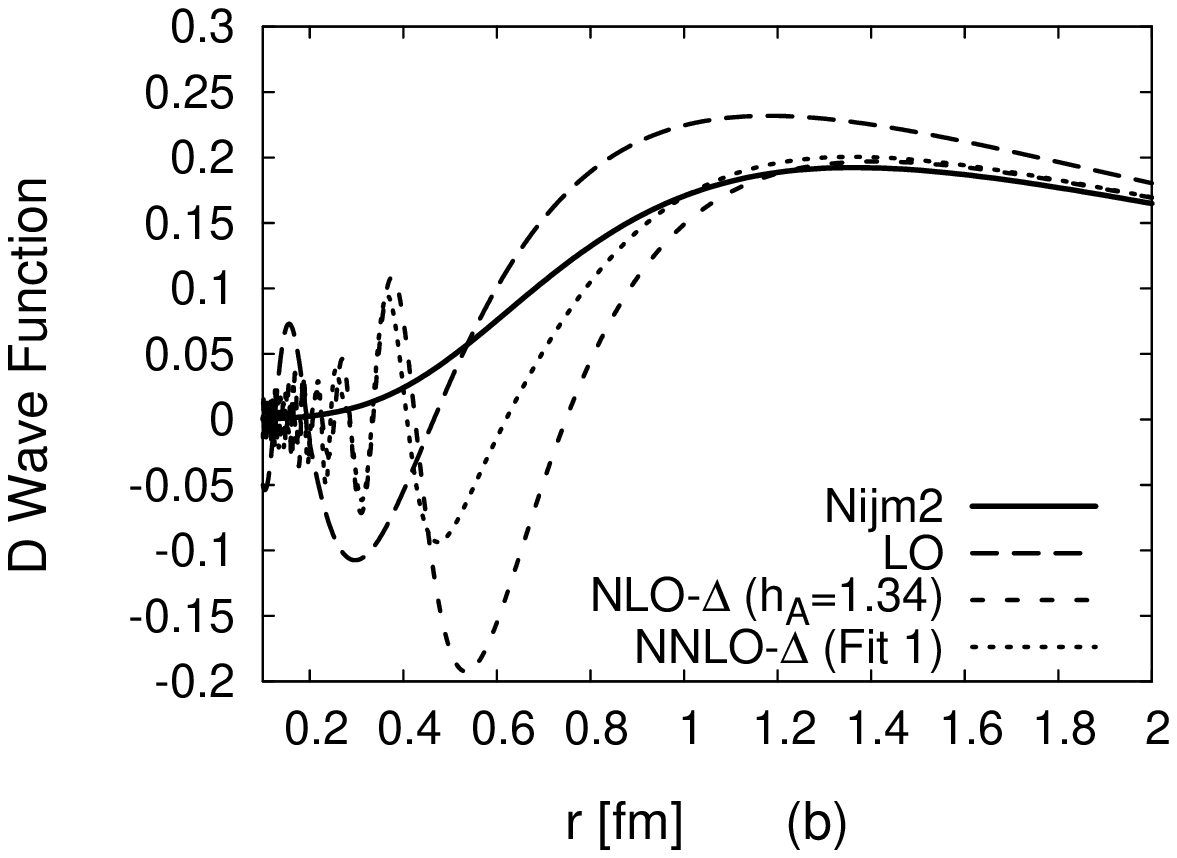,height=5cm,width=7cm}
\end{center}
\caption{Deuteron wave functions, u (left panel) and w (right panel),
as a function of the distance (in {\rm fm}) for the LO, NLO-$\Delta$ and
N2LO-$\Delta$ potentials, compared to the Nijmegen II wave
functions~\cite{Stoks:1994wp}. The asymptotic normalization $u(r) \to
e^{-\gamma r}$ has been adopted and the asymptotic D/S ratio is taken
$\eta = 0.0256 (4)$. The oscillations of the wave functions are related
to the presence of deeply bound states, but do not have any appreciable
effect on deuteron observables as they cannot be resolved in effective
field theory.
}
\label{fig:u+w_Delta}
\end{figure*} 

\section{The triplet channel} 
\label{sec:triplet} 

\subsection{Equations and boundary conditions}

The $^3S_1-{}^3D_1$ wave function in the pn c.m. system can be written as
\begin{eqnarray}
\Psi (\vec x) &=& \frac1{\sqrt{4\pi}\, r } \, \Big[ u(r) \sigma_p \cdot
\sigma_n \nonumber \\ &+& \frac{w(r)}{\sqrt{8}} \left( 3 \sigma_p
\cdot \hat x \, \sigma_n \cdot \hat x - \sigma_p \cdot \sigma_n \right)
\Big] \chi_{pn}^{s m_s} \, ,
\end{eqnarray} 
with total spin $s=1$ and $m_s=0,\pm 1$; $\sigma_p$ and $\sigma_n$ are 
the Pauli matrices for the proton and the neutron respectively. 
The functions $u(r)$ and $w(r)$ are the reduced S- and
D-wave components of the relative wave function respectively. 
They satisfy the coupled set of equations in the $^3S_1 - ^3D_1 $ channel
\begin{eqnarray}
-u '' (r) + U_{^3S_1} (r) u (r) + U_{E_1} (r) w (r) &=& k^2 u
 (r) \, ,\nonumber \\ -w '' (r) + U_{E_1} (r) u (r) + \left[U_{^3D_1}
 (r) + \frac{6}{r^2} \right] w (r) &=& k^2 w (r) \, , \nonumber
 \\
\label{eq:sch_coupled} 
\end{eqnarray}
with $ U_{^3S_1} (r)$, $U_{E_1} (r) $ and $U_{^3D_1} (r)$ the corresponding 
matrix elements of the coupled channel potential, which are
\begin{eqnarray}
U_{^3S_1} &=&  V_C -3 W_C  + V_S -3 W_S \, , \nonumber   
\\ 
U_{E_1} &=&  2 \sqrt{2} (V_T -3 W_T) \, ,   \nonumber 
\\ 
U_{^3D_1} &=& V_C - 3 W_C  + V_S - 3 W_S  - 2 V_T + 6 W_T \, . \nonumber \\    
\end{eqnarray} 
At short distances one has 
\begin{eqnarray}
U_{^3S_1} &\to&  \frac{M C_{6,^3S_1}}{r^6} \, , 
\nonumber \\ 
U_{E_1} &\to&  \frac{M C_{6,E_1}}{r^6}  \, , 
\nonumber \\ 
U_{^3D_1} &\to&   \frac{M C_{6,^3D_1}}{r^6}  \, ,
\nonumber \\    
\end{eqnarray} 
where the van der Waals coefficients are given by 
\begin{eqnarray}
C_{6,^3S_1} &=& C_{6,^3S_1}^{\rm NLO}  + C_{6,^3S_1}^{\rm N2LO} \, , 
\nonumber \\
C_{6,E_1} &=& C_{6,E_1}^{\rm NLO}  + C_{6,E_1}^{\rm N2LO} \, , 
\nonumber \\
C_{6,^3D_1} &=& C_{6,^3D_1}^{\rm NLO}  + C_{6,^3D_1}^{\rm N2LO}  \, .
\end{eqnarray} 
Their numerical values are listed in Table~\ref{tab:c_6-channels} for
Fits 1 and 2 of Ref.~\cite{Krebs:2007rh}. One can diagonalize the corresponding
matrix of van der Waals coefficients
\begin{eqnarray} 
\begin{pmatrix}
M C_{6,^3S_1} & M C_{6,E_1}\\ M C_{6,E_1}  & M C_{6,^3D_1}   
\end{pmatrix}  
&=&  
\begin{pmatrix}
\cos\theta & \sin\theta  \\ -\sin\theta &  \cos \theta   
\end{pmatrix}  
\begin{pmatrix}
-R_{+}^4  & 0  \\ 0 &  -R_{-}^4     
\end{pmatrix} \nonumber \\ &\times&   
\begin{pmatrix}
\cos\theta & -\sin\theta  \\ \sin\theta &  \cos \theta   
\end{pmatrix} \, ,
\end{eqnarray}  
where, according to Eq.~(\ref{eq:cos-theta}) the angle is 
\begin{eqnarray}
\cos \theta = \frac1{\sqrt{3}} \, ,
\end{eqnarray} 
i.e. $\theta = 54.7^o$ and common to both the NLO-$\Delta$ and
N2LO$-\Delta$ potentials. That means that the eigenvalues of the NLO-$\Delta$
and N2LO-$\Delta$ van der Waals matrices are additive, and therefore
can be summed up directly from Table~\ref{tab:c_6-channels}:  
\begin{eqnarray}
- R_{\pm}^4 &=&  - R_{\pm, {\rm NLO}}^4 - R_{\pm, {\rm N2LO}}^4 \,\, .
\end{eqnarray}
As we see from Table~\ref{tab:c_6-channels}, the NLO-$\Delta$ 
matrix is negative definite, but N2LO-$\Delta$ matrix is not so for Fit 1. 
However, what counts is the NLO-$\Delta$ + N2LO-$\Delta$
contribution which, as can be checked, is negative definite. 
In the diagonal basis one has at short distances 
\begin{eqnarray} 
\begin{pmatrix}
u \\ w 
\end{pmatrix}  
& \to & 
\begin{pmatrix}
\cos\theta & \sin\theta  \\ -\sin\theta &  \cos \theta   
\end{pmatrix}  
\begin{pmatrix}
v_+ \\ v_- 
\end{pmatrix} \, ,
\end{eqnarray}  
where the short distance eigen functions are 
%\begin{widetext} 
\begin{eqnarray} 
v_+ (r) &=& \left(\frac{r}{R_+}\right)^{\frac32} C_{+}
\sin\left[ \frac{1}{2} \frac{R_+^2}{r^2} + \varphi_+ \right] \, 
\nonumber \\ v_- (r)
&=& \left(\frac{r}{R_-}\right)^{\frac32}  C{-} \sin\left[
\frac{1}{2} \frac{R_-^2}{r^2} + \varphi_-\right] \, , \nonumber \\ 
\end{eqnarray}  
%\end{widetext} 
and $ \varphi_\pm $ are short distance phases which must be fixed
independently on the potential and $C_\pm $ suitable normalization
constants. Orthogonality of solutions of different energy requires
these phases to be energy independent. Following the procedure of
Ref.~\cite{Valderrama:2005wv} we fix them from deuteron physical
properties, namely the binding energy and asymptotic D/S ratio (see
below). Once these two quantities are fixed, scattering states can be 
completely determined by fixing in addition the scattering length of
the $^3S_1$ phase, and then imposing orthogonality to the deuteron state. 
This is equivalent to renormalizing with three counter-terms in momentum
space~\cite{Valderrama:2007ja}. Of course, more counter-terms could be
considered if orthogonality is given up by an energy dependent
boundary condition, as done above for the $^1S_0$ channel.

In the following two subsections we will consider the description of
the deuteron bound state and the scattering states and discuss our results.

\begin{table*}
%\begin{ruledtabular}
\begin{tabular}{|c|c|c|c|c|c|c|c|c|}
\hline
Set & $\gamma\,\,({\rm fm}^{-1})$ & $\eta$ & $A_S\,\,({\rm fm}^{-1/2})$ &
$r_m\,\,({\rm fm})$ & $Q_d ({\rm fm}^2)$ & $P_D$ (\%) &  $ \langle r^{-1} \rangle $ & $ \langle r^{-2} \rangle $ 
\\
\hline
LO & Input & 0.02633 & 0.8681(1) & 1.9351(5) & 0.2762(1) 
& 7.31(1) & 0.486 (1) & 0.434(3) \\
\hline\hline
NLO-$\Delta \negate$ &  Unbound  & --  & --   & -- & --  & --   & -- &-- 
\\ \hline
NLO-$\Delta$ ($h_A = 1.34$) & Input  &  Input  & 0.884(3) & 1.963(7)  & 0.274(9) & 5.9(4) &0.446(10)  & 0.29(2)  \\
NLO-$\Delta$ ($h_A = 1.05$) & Input  &  Input  & 0.84(4) & 1.86(8)  & 0.24(3) & 12(5) & 0.62(15)  & 0.8(4)  \\
\hline\hline
N2LO-$\Delta \negate$ (Set IV) & Input & Input & 0.884(4) & 1.967(6) & 0.276(3) & 8(1) &  0.447(5) & 0.284(8)\\
N2LO-$\Delta \negate$ ($\pi$N) & Input & Input & 0.896(2) & 1.990(3) & 0.282(5) & 6.1(8) &  0.4287(13) & 0.253(2) \\
\hline
N2LO-$\Delta$(Fit 1) &  Input  &  Input  & 0.892(2) &  1.980(4) & 0.279(5) & 5.9(9) & 0.4336(15) & 0.262(3)   \\
N2LO-$\Delta$(Fit 2) &  Input  &  Input  & 0.890(2) &  1.975(3) & 0.278(5) & 5.8(9) & 0.4470(15) & 0.268(2)   \\
\hline\hline
NijmII & 0.231605 & 0.02521 & 0.8845 & 1.9675 & 0.2707 & 5.635 & 0.4502 & 0.2868   \\
Reid93 & 0.231605 & 0.02514 & 0.8845 & 1.9686 & 0.2703 & 5.699 & 0.4515 & 0.2924    \\
\hline
Exp. 
& 0.231605 & 0.0256(4) & 0.8838(4) & 1.971(5) & 0.2860(15) 
& - & - & - \\
\hline
\end{tabular}
%\end{ruledtabular}
\caption{\label{tab:deut_prop} Deuteron properties for the OPE and TPE
potentials.  The computation is made by fixing $\gamma$ and $\eta$ to
their experimental values.  The errors quoted in both TPE computations
reflect the uncertainty in the non-potential parameters $\gamma$,
$\eta$ and $\alpha_0$ only.  For the OPE (LO) we take $g_{\pi
NN}=13.1(1)$. 
For the LEC's in the TPE calculation, by set IV we refer to the determination 
of Ref~\cite{Entem:2003ft} (see main text).
For the $\Delta$ case we use Fits 1 and 2 of
Ref.~\cite{Krebs:2007rh}. Fit 1 involves the SU(4) quark-model relation
$h_A= 3 g_A /(2 \sqrt{2}) = 1.34 $ for $g_A=1.26$. Fit 2 takes 
$h_A = 1.05$. The experimental values are taken from the following
references: $\eta$ from~\cite{PhysRevC.41.898}, 
$A_S$ from~\cite{PhysRevLett.60.1932}, $r_m$ from~\cite{PhysRevC.51.1127}
and $Q_d$ from~\cite{PhysRevA.20.381}
(see also Ref.~\cite{deSwart:1995ui} for a brief review).
}
\end{table*}

\subsection{The Deuteron}

In the case of negative energy we consider Eq.~(\ref{eq:sch_coupled}) with
\begin{eqnarray}
k^2 = -{\gamma^2} = - M\,B_d \, ,
\end{eqnarray} 
with $\gamma$ the deuteron wave number and $B_d$ the deuteron binding
energy. We solve Eq.~(\ref{eq:sch_coupled}) together with the
asymptotic condition at infinity
\begin{eqnarray}
u (r) &\to & A_S e^{-\gamma r} \, , \nonumber \\ 
w (r) & \to & A_D e^{-\gamma r} 
\left( 1 + \frac{3}{\gamma r} + \frac{3}{(\gamma r)^2} \right) \, ,
\label{eq:bcinfty_coupled} 
\end{eqnarray}
where  $A_S$ and $A_D$ are the s- and d-wave normalization factors. The 
asymptotic D/S ratio parameter $\eta$ is defined as $\eta = A_D / A_S$.

In order to obtain the regularized wave functions, we fix $\gamma$ and
$\eta$ to their experimental values (see table~\ref{tab:deut_prop}),
and obtain $u(r)$ and $w(r)$ by integrating Eq.~(\ref{eq:sch_coupled})
from $r \to \infty$ to $r = 0$ with~(\ref{eq:bcinfty_coupled}) as
boundary conditions.  $A_S$ can be later determined from
\begin{eqnarray}
\int_0^\infty dr \left[ u(r)^2 + w(r)^2 \right] = 1  \, , 
\label{eq:normalization} 
\end{eqnarray}
i.e., from demanding the deuteron normalization to be equal to unity.
The renormalized deuteron wave functions are depicted 
in Fig.~\ref{fig:u+w_Delta} at LO, NLO-$\Delta$ and N2LO-$\Delta$,
and compared to the Nijmegen II wave functions~\cite{Stoks:1994wp}. 
The asymptotic normalization $u \to e^{-\gamma r}$ has been adopted and
the asymptotic D/S ratio is taken to be $\eta = 0.0256$. 
One can see in Fig.~\ref{fig:u+w_Delta} the appearance of an increasing number
of oscillations in the wave function when the radius approaches zero. They
have no appreciable effect on the physics of the deuteron, as they happen at
very short distances. Therefore they have a very small effect on the 
computation of deuteron observables and cannot be resolved by external 
probes at the energies for which the effective theory description is valid.
This later point is shown explicitly in Ref.~\cite{Valderrama:2007ja} for
elastic electron-deuteron scattering. Perhaps the most
remarkable aspect of the present calculation is the convergence of the
proposed scheme when the $\Delta$-resonance is considered, which of
course implies that the deuteron is bound at all computed orders 
of the potential. 
This is in contrast with the delta-less theory, where at NLO-$\Delta \negate$
the deuteron became unbound (as discussed extensively
in Ref.~\cite{Valderrama:2005wv}).

Here, we also compute the matter radius, which reads,
\begin{eqnarray}
r_m^2 = \frac{\langle r^2 \rangle}{4} = \frac14 \int_0^\infty r^2 ( u(r)^2 +
w(r)^2 ) dr \, , 
\end{eqnarray} 
the quadrupole moment (without meson exchange currents) 
\begin{eqnarray}
Q_d  = \frac1{20} \int_0^\infty r^2 w(r) ( 2\sqrt{2} u(r)-w(r) ) dr  \, , 
\end{eqnarray} 
the $D$-state probability
\begin{eqnarray}
P_D = \int_0^\infty w(r)^2  dr \, ,
\end{eqnarray} 
and the inverse moments of the radius
\begin{eqnarray}
\langle r^{-n} \rangle = \int_0^\infty r^{-n} ( u(r)^2 + w(r)^2 ) dr
\, ,
\end{eqnarray} 
which, as is well known, appear in the multiple scattering expansion of the 
$\pi$-deuteron scattering length. Some results for the inverse moments
in $\Delta$-less effective field theory can be found in 
Refs.~\cite{Valderrama:2006np, RuizArriola:2006xs}.
In Tab.~\ref{tab:deut_prop} we show our results in a variety of situations.
In general, the results for NLO-$\Delta$ and N2LO-$\Delta$ are in agreement
with the experimental value of the deuteron observables, with the exception
of the quadrupole moment, which presents a discrepancy of $0.01\,{\rm fm}^2$.
The reason for it lies in meson exchange current contributions to the 
quadrupole moment, which were estimated for ChPT 
in Ref.~\cite{Phillips:2003jz},
and are of the order of the difference of our results with respect to
the experimental value.
When comparing the $\Delta$-full computations with the $\Delta$-less ones,
one can notice, in the first place, that a renormalized result exists for 
NLO-$\Delta$, unlike the NLO-$\Delta \negate$ case
(as was explained in more detail in Ref.~\cite{Valderrama:2005wv}).
This supports the better convergence properties of effective theory
when including the $\Delta$ degree of freedom.
It can also be noticed that the D-wave probability $P_D$, although not an 
observable, it is better described in N2LO-$\Delta$ than in 
N2LO-$\Delta \negate$, when compared with the results coming from the
phenomenological potentials~\cite{deSwart:1995ui}. 
The consequence is that the magnetic moment
of the deuteron would be better described in the $\Delta$-full theory
than in the $\Delta$-less theory, although no actual computation 
has been made on that respect in the present paper.
The reason for that is that in a non-relativistic framework, the deuteron
magnetic moment depends solely on $P_D$~\cite{Gilman:2001yh}.
Before comparing the results for other observables, one important comment
must be made: one should only directly compare the N2LO-$\Delta$ 
results with the N2LO-$\Delta \negate$ ($\pi$N) ones. 
Comparison with N2LO-$\Delta \negate$ set IV should be done with care,
as for this case the LECs are fitted to reproduce NN data~\cite{Entem:2003ft}.
This is why they look slightly better than the other results of 
Table~\ref{tab:deut_prop}. 
Therefore, the comparison of N2LO-$\Delta$ with N2LO-$\Delta \negate$ 
($\pi$N) results implies in particular that the inclusion
of $\Delta$ enhances the compatibility between $\pi N$ and $NN$ scattering,
as one would expect within the EFT philosophy.

\begin{figure*}[tbc]
\begin{center}
\epsfig{figure=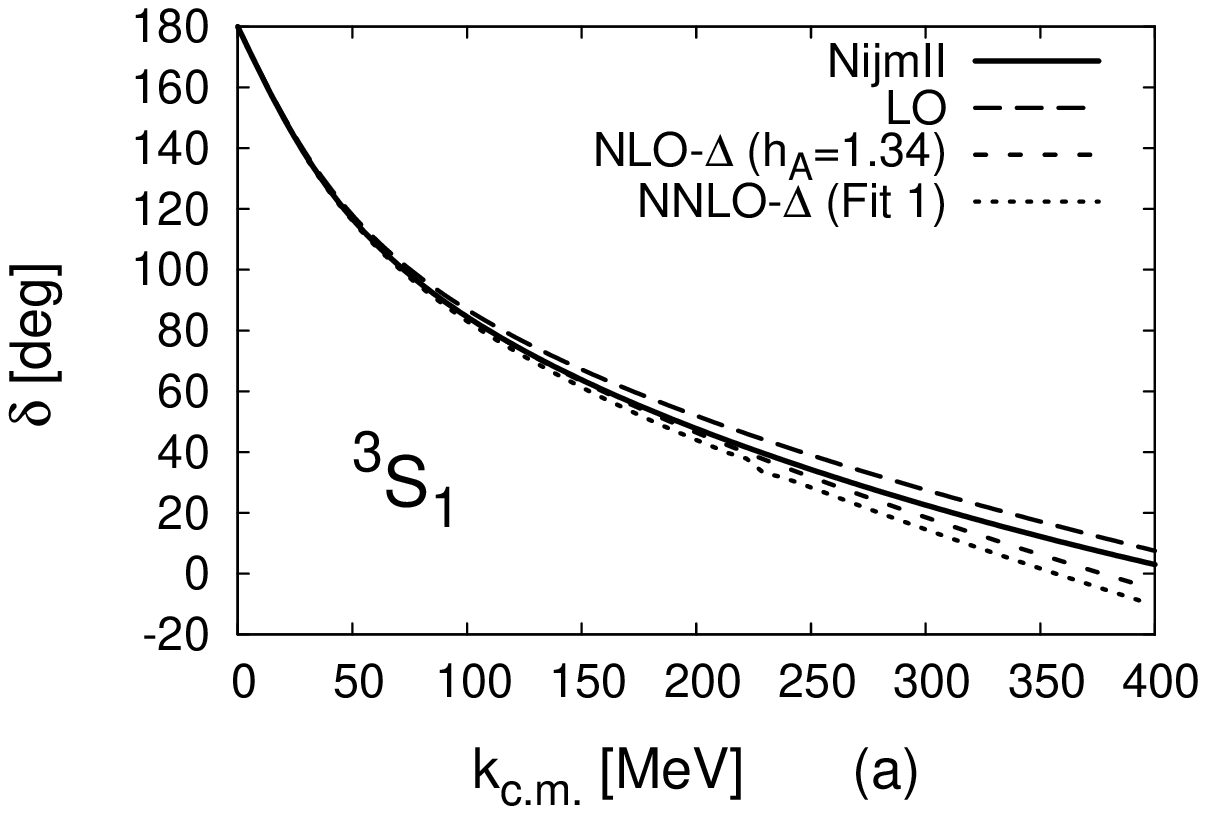, height=6cm, width=5.5cm}
\epsfig{figure=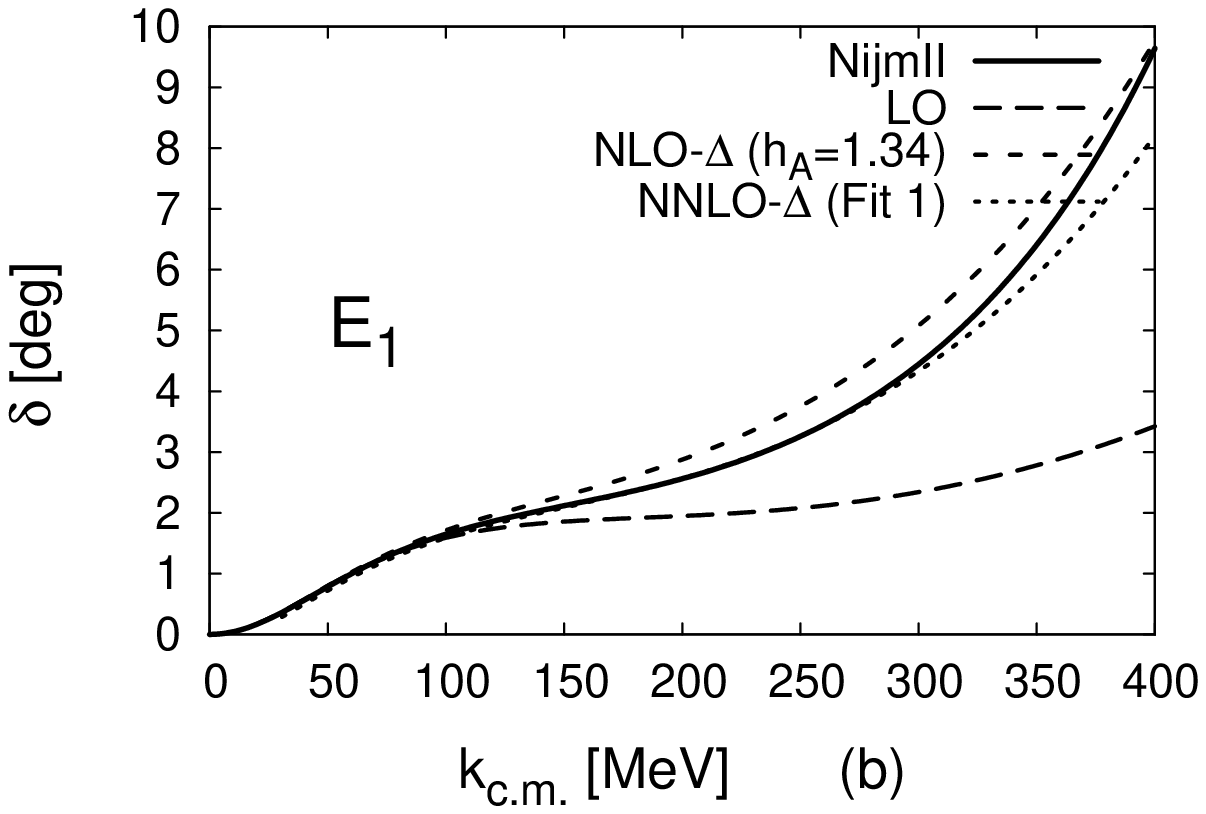, height=6cm, width=5.5cm}
\epsfig{figure=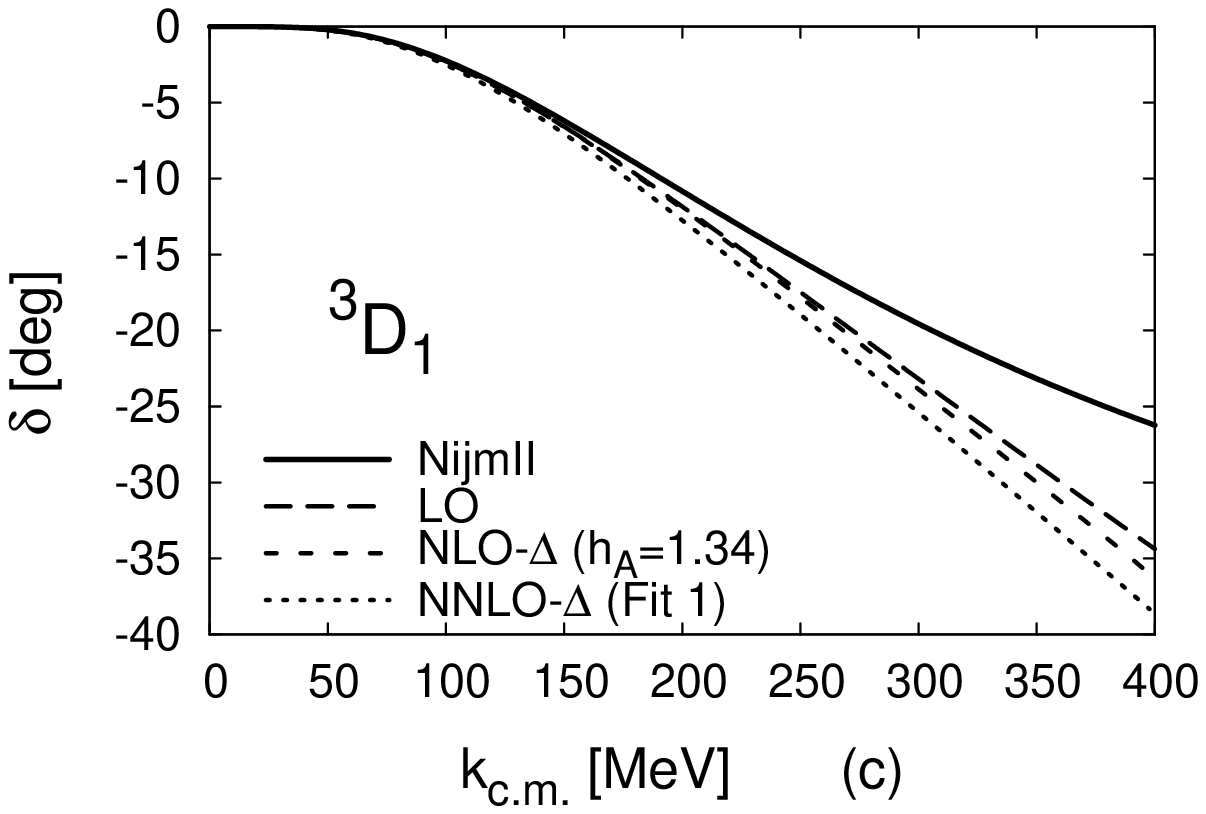, height=6cm, width=5.5cm}
\end{center}
\caption{np spin triplet (eigen) phase shifts for total
angular momentum $j=1$ as a function of the c.m. momentum, 
compared to the Nijmegen II potential phase shifts~\cite{Stoks:1994wp}.
}
\label{fig:phase-triplet-Delta}
\end{figure*}

\subsection{Phase Shifts}

Finally, in the case of positive energy we consider
Eq.~(\ref{eq:sch_coupled}) with
\begin{eqnarray}
E_{c.m.} = \frac{k^2}{M} 
\end{eqnarray} 
where $k$ is the corresponding c.m. momentum.  
We solve Eq.~(\ref{eq:sch_coupled})
for the two linear independent scattering states, which are usually 
labelled as the $\alpha$ and $\beta$ states. 
They are defined by the asymptotic normalization
\begin{eqnarray}
u_{k,\alpha} (r) &\to & {\cos \epsilon}\Big( \hat
j_0 (kr) \cot \delta_1 - \hat y_0 (kr) \Big) \, , \nonumber \\ w_{k,\alpha}
(r) &\to & {\sin \epsilon} \Big( \hat j_2 (kr)\cot \delta_1  -
\hat y_2(kr)\Big) \, , \nonumber \\ \\  
u_{k,\beta} (r) & \to & -{\sin \epsilon}\Big( \hat j_0 (kr) \cot \delta_2 - 
\hat y_0 (kr)\Big) \, ,  \nonumber \\ 
w_{k,\beta} (r) &\to & {\cos \epsilon}\Big( \hat
j_2 (kr) \cot \delta_2 - \hat y_2(kr) \Big) \, , \nonumber \\ 
\label{eq:phase_triplet}
\end{eqnarray} 
where $ \hat j_l (x) = x j_l (x) $ and $ \hat y_l (x) = x y_l (x) $
are the reduced spherical Bessel functions and $\delta_1$ and
$\delta_2$ are the eigen-phases in the $^3S_1$ and $^3D_1$ channels;
$\epsilon$ is the mixing angle $E_1$. 
The orthogonality constraints between the deuteron and scattering states
generate the following boundary conditions
\begin{eqnarray}
u_\gamma u_{k,\alpha}' - u_\gamma' u_{k,\alpha} + w_\gamma
w_{k,\alpha}' - w_\gamma ' u_{k,\alpha} \Big|_{r=r_c} &=& 0 \, , \nonumber \\
u_\gamma u_{k,\beta}' - u_\gamma' u_{k,\beta} + w_\gamma w_{k,\beta}'
- w_\gamma ' u_{k,\beta} \Big|_{r=r_c} &=& 0 \, . \nonumber \\
\label{eq:orth_triplet_k} 
\end{eqnarray} 
The use of the superposition principle for the $\alpha$ and $\beta$
scattering states, plus the deuteron wave functions, allows to deduce 
the corresponding $^3S_1-{}^3D_1$ eigen phase-shifts. 
The results are depicted in Fig.~\ref{fig:phase-triplet-Delta} at LO, 
NLO-$\Delta$ and N2LO-$\Delta$ compared to the Nijmegen II 
potential results~\cite{Stoks:1994wp}.
We observe a clear improvement in the threshold region, and quite remarkably
for the $E_1$ mixing phase. However, there is no improvement in the $^3D_1$ 
phase-shift.

\section{Comparison with the Delta-less theory}
\label{sec:comparison}

In our previous work~\cite{Valderrama:2005wv}, the renormalization of
Delta-less theory was analyzed. As discussed above, it was found that
at NLO-$\Delta \negate$ there was no deuteron bound state 
if the cut-off was removed. 
The reason was due to the short distance $\sim g_A^4 / ( f_\pi^4 r^5)$ 
repulsive singular character of the potential in the $^3S_1-{}^3D_1$ channel. 
However, at N2LO-$\Delta \negate$ three counter-terms where needed due to the 
short distance attractive singular character of the potential. 
In fact the agreement with more sophisticated calculations~\cite{Entem:2003ft} 
was remarkable. 
More recently, the quality of these chiral wave functions has been tested in
electron-deuteron scattering~\cite{Valderrama:2007ja} using LO
currents, with an amazingly good agreement up to momentum transfer of 
$q = 1 {\rm GeV}$.  
It is interesting to compare the results of the Delta-less 
theory~\cite{Valderrama:2005wv} with the ones found here after 
inclusion of the $\Delta$ in the potential. 
The first aspect to note is that the NLO-$\Delta$ potential has an 
attractive $\sim g_A^4 / (\Delta f_\pi^4 r^6)$ short distance singularity, 
a feature kept at the N2LO-$\Delta$, however with different scales. 

\begin{figure*}[]
\begin{center}
\epsfig{figure=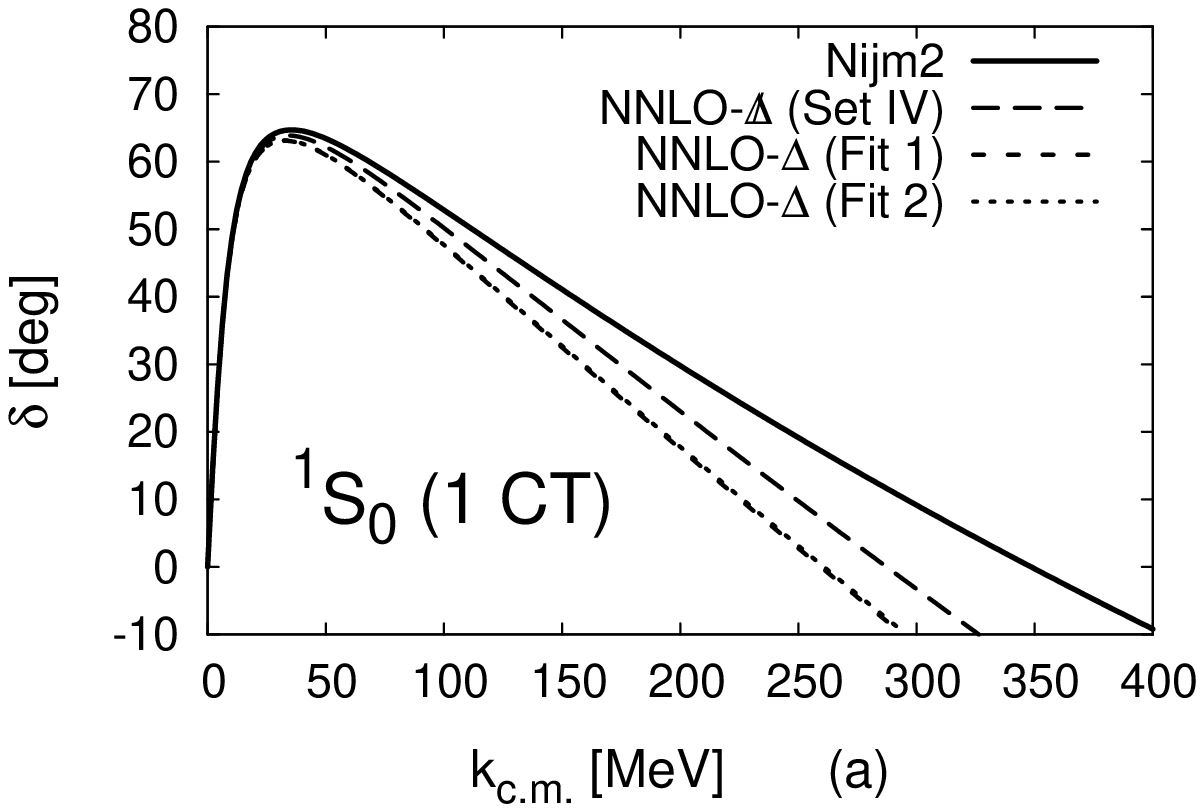, height=6.5cm, width=6.5cm}
\epsfig{figure=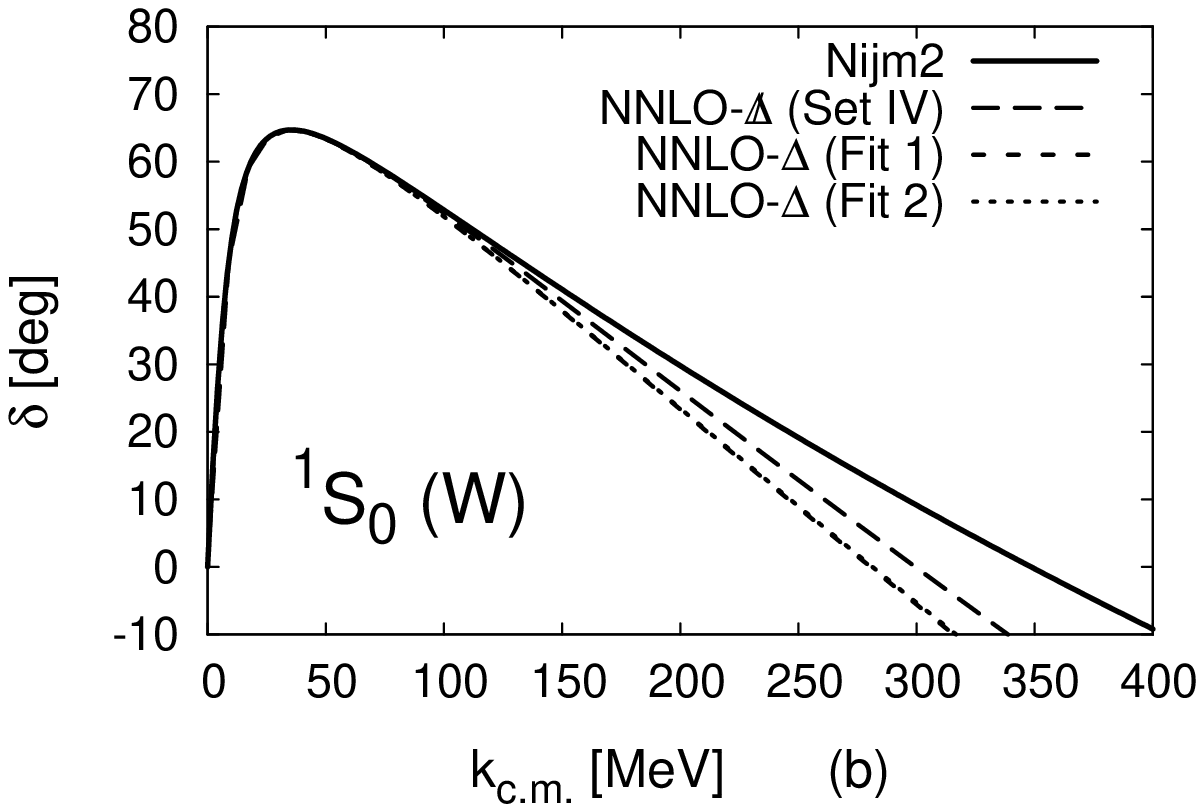, height=6.5cm, width=6.5cm}
\end{center}
\caption{The np spin singlet phase shifts as a function of the c.m. momentum
for the N2LO-$\Delta \negate$ and N2LO-$\Delta$ potentials, 
with one counter-term (left panel) and two counter-terms (right panel),
compared to the Nijmegen II potential results~\cite{Stoks:1994wp}.}
\label{fig:phase-singlet-Delta}
\end{figure*}

In Fig.~\ref{fig:phase-singlet-Delta} the np $^1S_0$ singlet phase shifts
are depicted as a function of the c.m. momentum for the N2LO-$\Delta$
and N2LO-$\Delta \negate$ compared to the Nijmegen II potential
phase shifts~\cite{Stoks:1994wp}. 
As we see, with only one counter-term, i.e. fixing the scattering length 
$\alpha_0$, the result is slightly worsened when the $\Delta$ is included. 
This is consistent with the change in the effective range reported in
Table~\ref{tab:table_singlet}. Of course, if $r_0$ is fixed as an
additional renormalization condition, there is an improvement in the
low energy region but the difference between including or not $\Delta$
degrees of freedom is hardly visible.

\begin{figure*}[]
\begin{center}
\epsfig{figure=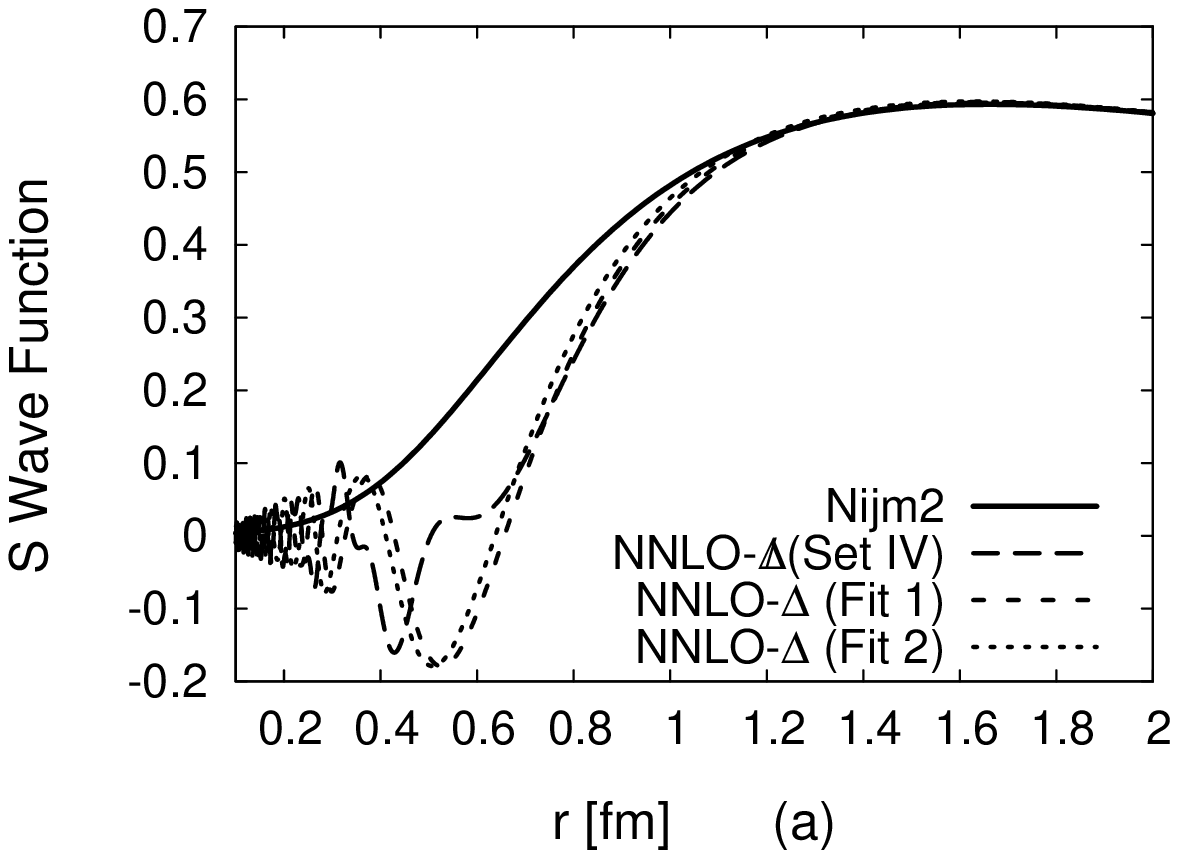,height=5cm,width=7cm}
\epsfig{figure=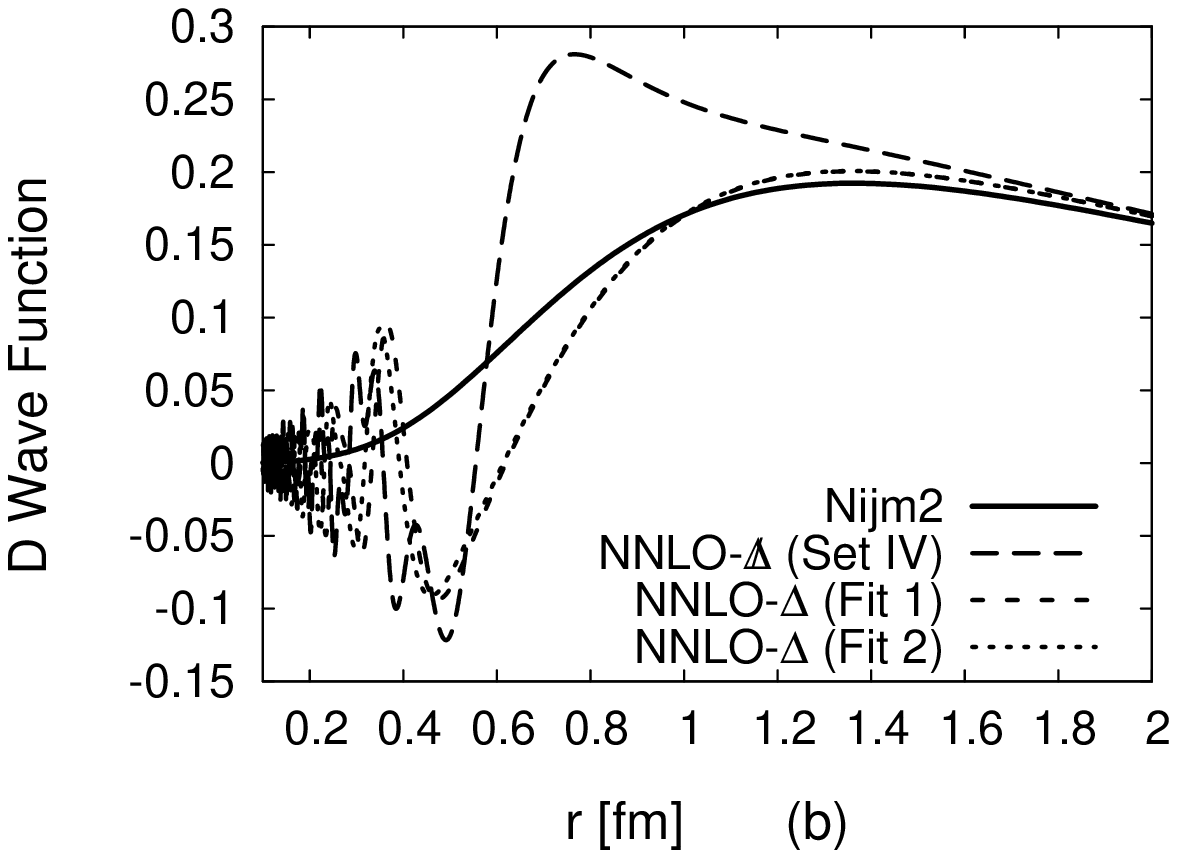,height=5cm,width=7cm}
\end{center}
\caption{Deuteron wave functions, u (left panel) and w (right panel), as a
function of the radius (in {\rm fm}) for the N2LO-$\Delta \negate$ potential
and the NLO and N2LO-$\Delta$ potentials, compared to the Nijmegen II wave
functions~\cite{Stoks:1994wp} . The asymptotic normalization $u \to
e^{-\gamma r}$ has been adopted and the value $\eta = 0.0256 (4)$
is taken for the asymptotic D/S ratio.}
\label{fig:u+w_TPE}
\end{figure*}

The situation for the deuteron wave functions is slightly different.
As we see in Fig.~\ref{fig:u+w_TPE}, the present N2LO-$\Delta$ deuteron
wave functions resemble slightly better the Nijmegen II ones
\cite{Stoks:1993tb,Stoks:1994wp}. 
In particular, the D-wave becomes smaller when one goes from our 
previous TPE ones~\cite{Valderrama:2005wv}.

\begin{figure*}[]
\begin{center}
\epsfig{figure=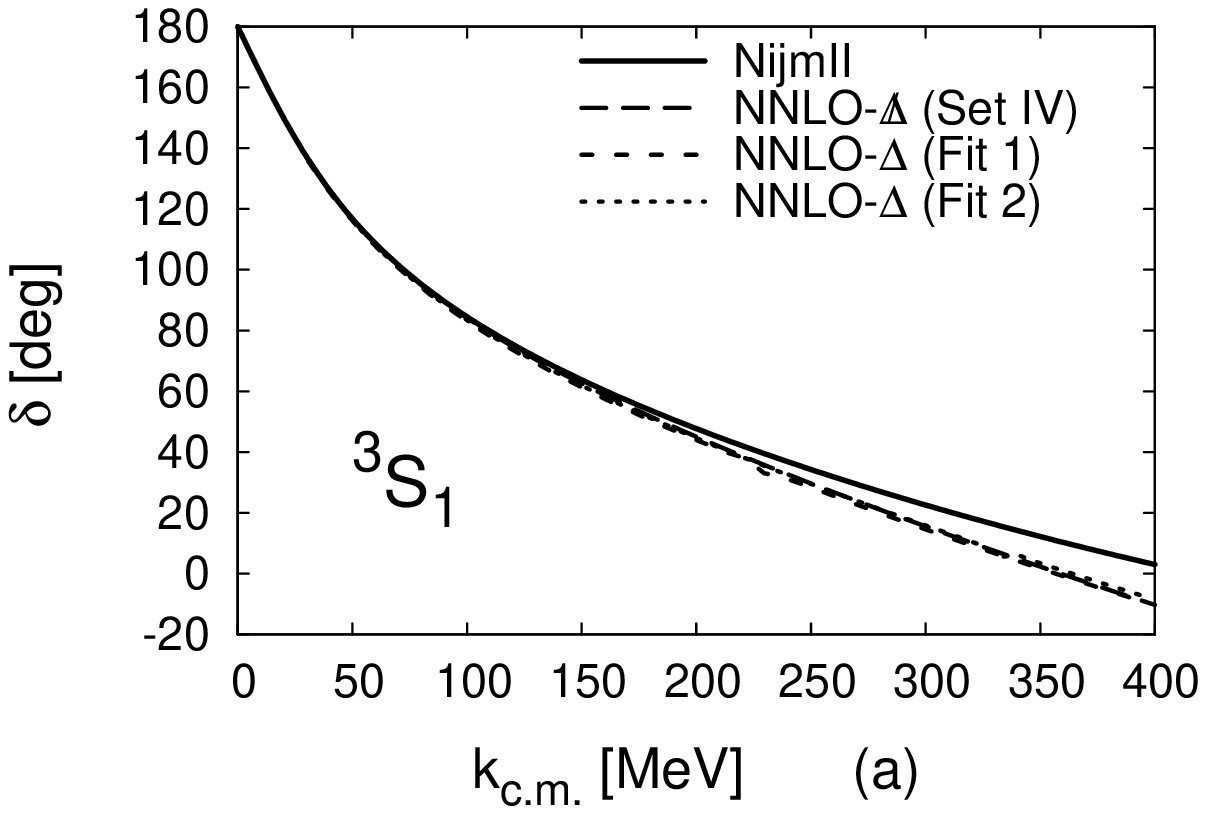, height=6cm, width=5.5cm}
\epsfig{figure=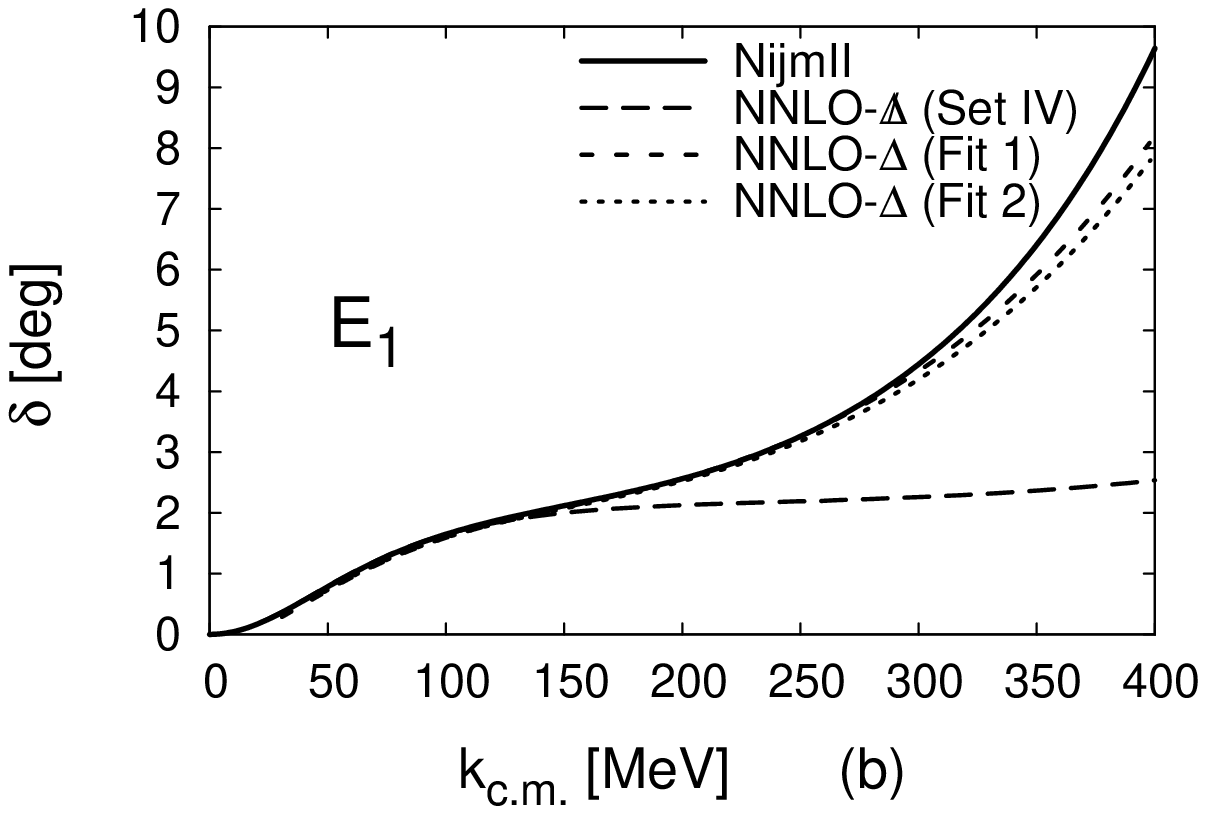, height=6cm, width=5.5cm}
\epsfig{figure=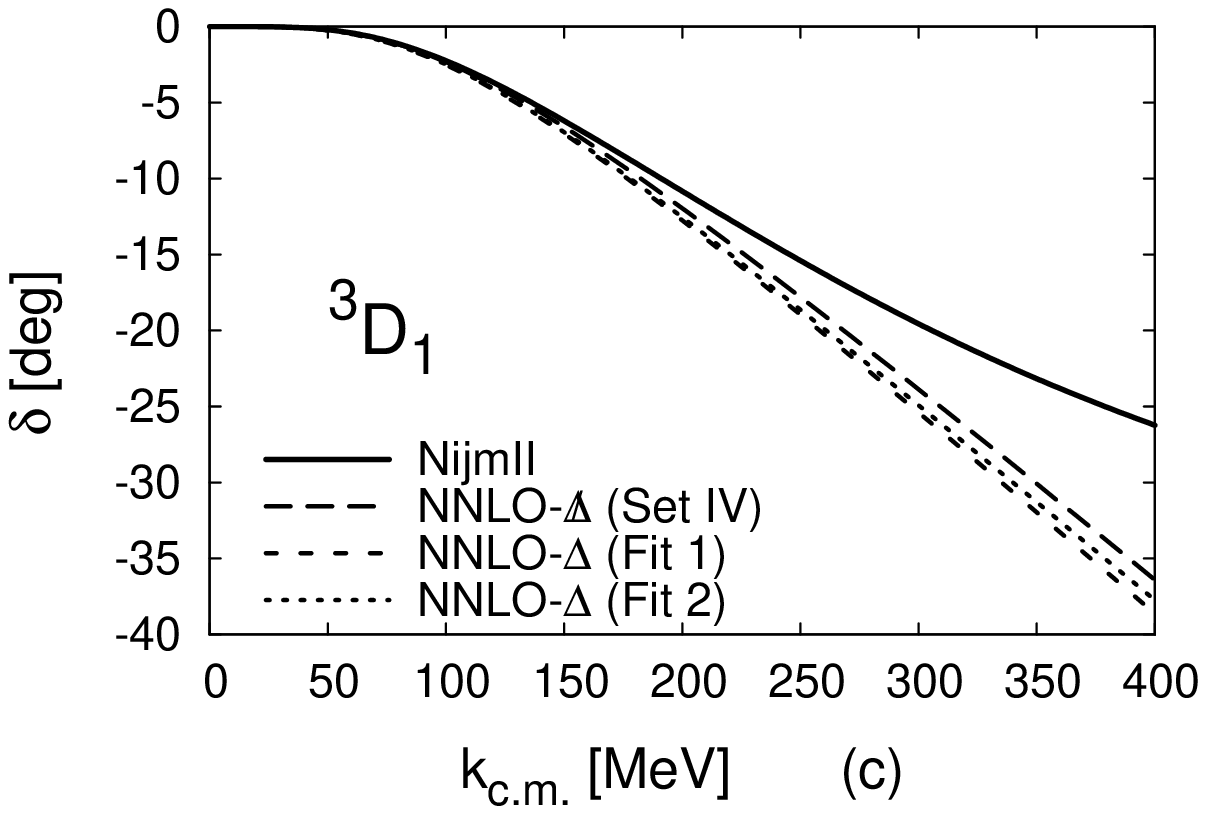, height=6cm, width=5.5cm}
\end{center}
\caption{The np spin triplet (eigen) phase shifts for the total
angular momentum $j=1$ as a function of the c.m. momentum for the
N2LO-$\Delta$ and N2LO-$\Delta \negate$ potentials compared to the 
Nijmegen II potential results~\cite{Stoks:1993tb,Stoks:1994wp}.}
\label{fig:phase-triplet-TPE}
\end{figure*}

Finally, in Fig.~\ref{fig:phase-triplet-TPE} we show the np 
eigen phase shifts in the $^3S_1-{}^3D_1$ channel as a function of 
the c.m. momentum for the N2LO-$\Delta$ and N2LO-$\Delta \negate$ compared to
the Nijmegen II potential results~\cite{Stoks:1994wp}.
The noticeable improvement in the $E_1$ phase is in agreement with 
the smaller $D$-wave deuteron wave function.


\section{Scheme dependence and cut-offs}
\label{sec:spectral} 

\subsection{The relevant scales}

As we see our scheme including the $\Delta$ does not reproduce the
$^1S_0$ phase shift for momenta larger than the pion mass even if the
effective range is adjusted to its experimental value. This trend has
also been observed in other renormalized calculations including TPE
effects and 1/M corrections in a Heavy Baryon 
formalism~\cite{Valderrama:2005wv}, using a relativistic 
potential~\cite{Higa:2007gz} or including N3LO
contributions~\cite{Entem:2007jg}. 
Within the chiral approach to NN interactions the main candidates for 
explaining such a disagreement would be $3\pi$-exchange or 
relativistic corrections.

\begin{figure*}[tbc]
\begin{center}
\epsfig{figure=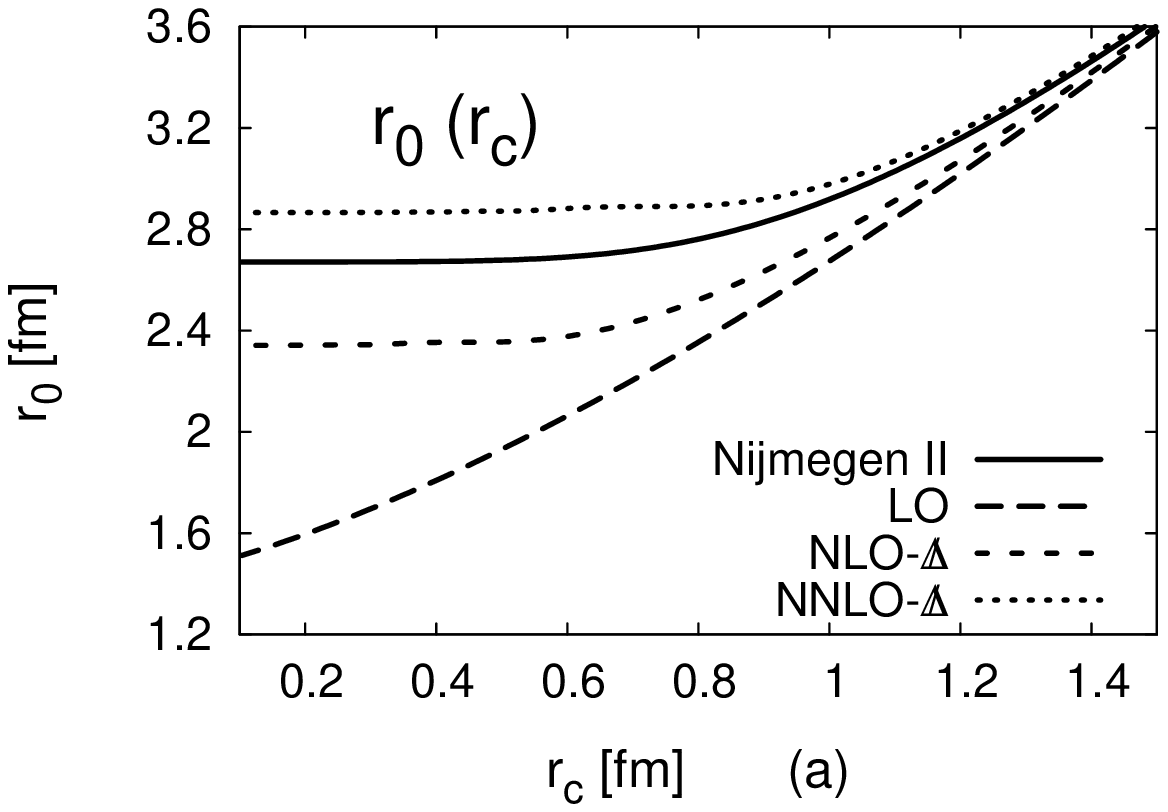,height=6.5cm,width=6.5cm}
\epsfig{figure=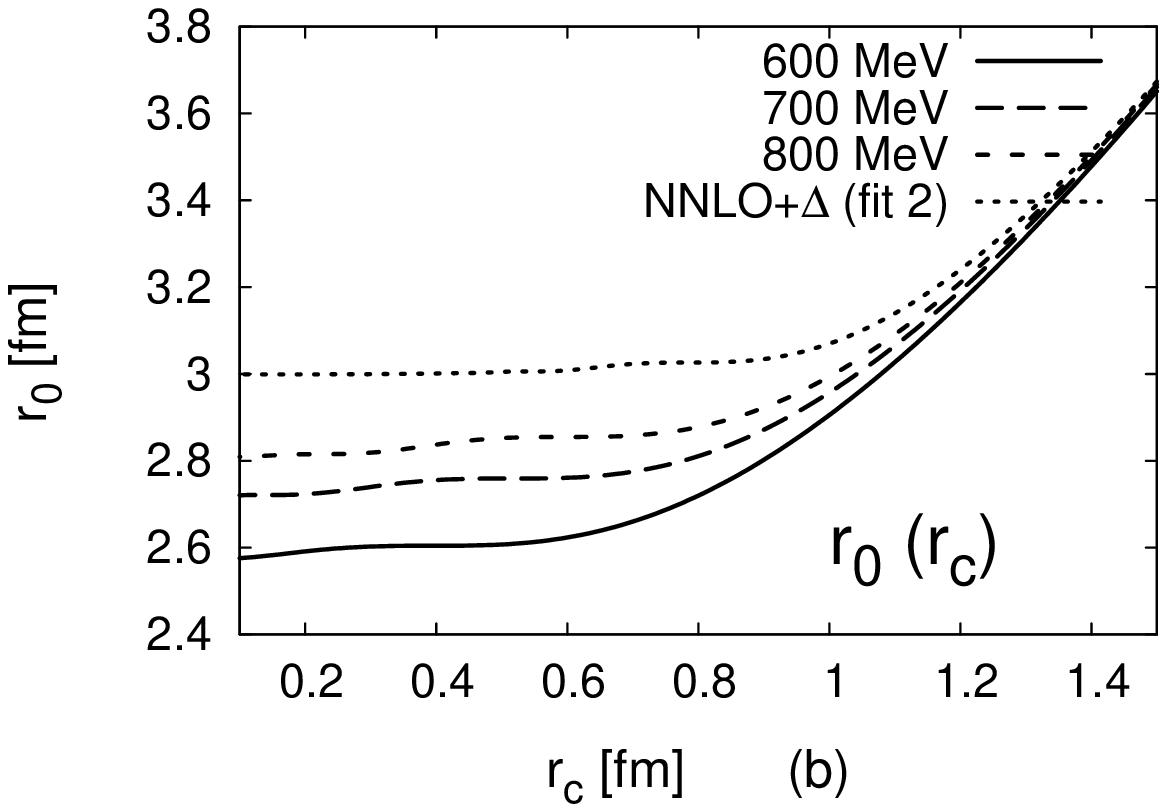,height=6.5cm,width=6.5cm}
\end{center}
\caption{$^1S_0$ effective range (in fm) $r_0(r_c) =
2\,\left(\int_0^{\infty}(1-r/\alpha_0)^2 \,dr
-\int_{r_c}^{\infty}u_0^2\,dr \right) $, with $\alpha_0 = -23.74 {\rm
fm}$ as a function of the short distance cut-off $r_c$ (in fm). We
compare the (Heavy Baryon) Delta-less theory at LO, NLO and N2LO and
the Nijmegen II potential~\cite{Stoks:1994wp} (left panel) with the
spectral-regularized N2LO-$\Delta$ potential for different values of
the spectral cut-off $\tilde \Lambda$ (right panel).}
\label{fig:r0-rc-spectral}
\end{figure*}

It is interesting at this point to ask which are the relevant scales
which build the full strength of the phase shifts. As already found
in~\cite{PavonValderrama:2005gu,Valderrama:2005wv,PavonValderrama:2005uj,
Entem:2007jg,Higa:2007gz} this is about $0.4-0.5\,{\rm fm}$ for
Delta-less calculations. For illustration purposes, the short distance
cut-off dependence is shown in Fig.~\ref{fig:r0-rc-spectral} for the
effective range in the case of renormalization with one counter-term in
the Heavy Baryon Delta-less theory to LO, NLO-$\Delta \negate$ and 
N2LO-$\Delta \negate$ where also the
Nijmegen II potential~\cite{Stoks:1994wp} is considered. 
As we see, and regardless on the full renormalized effective range value, the
Nijmegen II saturating scale is in between NLO-$\Delta \negate$ and
N2LO-$\Delta \negate$, a not unreasonable result. Likewise the
N2LO-$\Delta$ effective range displays a similar approach to the
renormalized result.

These scales are comparable to the nucleon size but also to the range
where $3\pi$ exchange starts contributing since it behaves as 
$e^{- 3 m_\pi r}$ at long distances.
According to the results of 
Kaiser~\cite{Kaiser:1999ff,Kaiser:1999jg,Kaiser:2001dm} (and in the
absence of $\Delta$) the potential is attractive, and using 
dimensional estimates for the $3\pi$-exchange potential, it behaves as
$g_A^6/(f_\pi^6 r^7)$ at short distances indicating a stronger
singularity and therefore a stronger short distance $ u(r) \sim
r^\frac74$ suppression as well. Thus, we expect that including these
effects would only slightly worsen the results by reducing the
phase-shift. 

Another interesting observation is that by taking larger coordinate
space cut-offs the phase shift is not improved, as in our
regularization procedure it converges from below, so in this case the
best possible cut-off is $r_c \to 0$~\footnote{Improving or not the
  phase shifts by using a finite cut-off while keeping the same
  renormalization conditions is, in fact, a regularization scheme
  dependent feature, and hence a further good reason to remove the
  cut-off. This choice however does not resolve the representation
  dependence of the potential on the choice of the pion field.
  Therefore, the cut-off less solution will obviously depend on the
  previous choice.}.  In this context, one can identify two kind of
finite cut-off sources of errors: the first one is related with the
explicit form of the regulator employed in the computations, and the
second one with the actual size of the finite cut-off.  This second
source of errors is the only one which is usually assessed in most
effective field theory works, while the role played by the regulator
is commonly ignored, probably resulting in an underestimation of the
errors.  While in our present renormalization scheme the most sensible
thing to do in the $^1S_0$ singlet channel is to completely remove the
cut-off, this may be not be the case in other regularization schemes.
Although this could be in fact considered as a good motivation for
keeping a finite cut-off or introducing form factors, we think that
this kind of procedure is difficult to justify from the EFT viewpoint.

The influence of relativistic effects on the results is less obvious,
since they generate either energy dependence or non-localities, but
experience in renormalizing relativistic potentials indicates that
they are not crucial, at least in the $^1S_0$
channel~\cite{Higa:2007gz}, and are saturated by scales larger than
the nucleon Compton wavelength, i.e. well above a possible influence
from the $N \bar N$ channel. 

Thus, we see that in all these calculations including TPE effects the
saturating scales are of the order of $0.5\,{\rm fm}$, regardless on
detailed issues, in particular including or not the $\Delta$. This is
comparable to other scales, $2 /m_\rho \sim 0.51\,{\rm fm}$, which may
equally represent vector meson exchange or nucleon size effects.
Therefore the discrepancy between renormalized phase shifts and
phenomenological ones should be considered a real one, in the sense
that finite size effects may be important physically.  Moreover, it
should be kept in mind that once the short distance cut-off becomes
smaller than the nucleon size, the effect of the counter-terms is not
enough to reproduce intermediate energy phase shifts, meaning
that they cannot mimic the finite nucleon size or other short-distance
physical effects which may be responsible of the observed discrepancy.

\begin{table*}
%\begin{ruledtabular}
\begin{tabular}{|c|c|c|c|c|c|c|c|c|}
\hline
Set & $\gamma\,\,({\rm fm}^{-1})$ & $\eta$ & $A_S\,\,({\rm fm}^{-1/2})$ &
$r_m\,\,({\rm fm})$ & $Q_d ({\rm fm}^2)$ & $P_D$ (\%) &  $ \langle r^{-1} \rangle $ & $ \langle r^{-2} \rangle $ 
\\
\hline
LO & Input & 0.02633 & 0.8681(1) & 1.9351(5) & 0.2762(1) 
& 7.31(1) & 0.486 (1) & 0.434(3) \\
\hline\hline
NLO-$\Delta \negate$ &  Unbound  & --  & --   & -- & --  & --   & -- &-- 
 \\ \hline
${\tilde \Lambda} = 700\,{\rm MeV}$ & Input  &  0.02669  & 0.851(7) & 1.894(15)  & 0.270(4) & 8.9(1.0) & 0.6(2)  & 1.8(1.5) \\
$r_c = 0.5\,{\rm fm}$ & Input  &  Input  & 0.83(2) & 1.86(6)  & 0.24(2) & 8(3) & 0.56(7)  & 0.51(15) \\
\hline \hline
NLO-$\Delta$ ($h_A = 1.34$) & Input  &  Input  & 0.884(3) & 1.963(7)  & 0.274(9) & 5.9(4) &0.446(10)  & 0.29(2)  \\ \hline
${\tilde \Lambda} = 700\,{\rm MeV}$ & Input  &  0.02637  & 0.8720(10) & 1.938(2)  & 0.2781(6) & 7.31(12) & 0.48(2)  & 0.5(2) \\
$r_c = 0.5\,{\rm fm}$ & Input  &  Input  & 0.867(13) & 1.93(3)  & 0.263(14) & 6.6(1.2) & 0.48(4)  & 0.35(8) \\
\hline\hline
NLO-$\Delta$ ($h_A = 1.05$) & Input  &  Input  & 0.84(4) & 1.86(8)  & 0.24(3) & 12(5) & 0.62(15)  & 0.8(4)  \\ \hline
${\tilde \Lambda} = 700\,{\rm MeV}$ & Input  &  0.02651  & 0.864(2) & 1.922(5)  & 0.2755(14) & 7.8(3) & 0.52(5)  & 0.7(4) \\
$r_c = 0.5\,{\rm fm}$ & Input  &  Input  & 0.854(17) & 1.90(4)  & 0.26(2) & 7(2) & 0.51(5)  & 0.41(11) \\
\hline\hline
N2LO-$\Delta \negate$ (Set IV) & Input & Input & 0.884(4) & 1.967(6) & 0.276(3) & 8(1) &  0.447(5) & 0.284(8)\\\hline
${\tilde \Lambda} = 700\,{\rm MeV}$ &  Input  &  0.02504  & 0.877(2) &  1.946(5) & 0.264(2) & 5.6(7) & 0.48(2) & 0.42(6) \\
$r_c = 0.5\,{\rm fm}$ &  Input  &  Input  & 0.875(8) &  1.942(15) & 0.270(2) & 8(2) & 0.46(2) & 0.27(6) \\
\hline \hline
N2LO-$\Delta \negate$ ($\pi$N)
& Input & Input & 0.896(2) & 1.990(3) & 0.282(5) & 6.1(8) &  0.4287(13) & 0.253(2) 
\\ \hline
${\tilde \Lambda} = 700\,{\rm MeV}$ &  Input  &  0.02590  & 0.8801(2) &  1.9540(5) & 0.27659(14) & 6.580(6) & 0.463(4) & 0.35(3) \\
$r_c = 0.5\,{\rm fm}$ &  Input  &  Input  & 0.883(4) &  1.956(6) & 0.274(9) & 5.9(8) & 0.447(13) & 0.28(2) \\
\hline\hline
N2LO-$\Delta$(Fit 1) &  Input  &  Input  & 0.892(2) &  1.980(4) & 0.279(5) & 5.9(9) & 0.4336(15) & 0.262(3)   \\
${\tilde \Lambda} = 700\,{\rm MeV}$ & Input  &  0.02603  & 0.8783(3) & 1.9507(6)  & 0.2774(2) & 6.77(3) & 0.465(4)  & 0.36(4) \\
$r_c = 0.5\,{\rm fm}$ & Input  &  Input  & 0.880(6) & 1.954(14)  & 0.272(10) & 5.9(6) & 0.45(2)  & 0.29(4) \\ \hline\hline
N2LO-$\Delta$(Fit 2) &  Input  &  Input  & 0.890(2) &  1.975(3) & 0.278(5) & 5.8(9) & 0.4470(15) & 0.268(2)   \\
${\tilde \Lambda} = 700\,{\rm MeV}$ &  Input  &  0.02606  & 0.8769(4) &  1.9479(7) & 0.2770(2) & 6.83(3) & 0.467(5) & 0.37(4) \\
$r_c = 0.5\,{\rm fm}$ &  Input  &  Input  & 0.878(6) &  1.950(15) & 0.271(10) & 5.9(6) & 0.46(2) & 0.30(4) \\
\hline\hline
NijmII & 0.231605 & 0.02521 & 0.8845 & 1.9675 &  5.635 & 0.2707 & 0.4502 & 0.2868   \\
Reid93 & 0.231605 & 0.02514 & 0.8845 & 1.9686 & 0.2703 & 5.699 & 0.4515 & 0.2924    \\
\hline
Exp. 
& 0.231605 & 0.0256(4) & 0.8838(4) & 1.971(5) & 0.2860(15) 
& - & - & - \\
\hline
\end{tabular}
%\end{ruledtabular}
\caption{\label{tab:deut_prop_sr} Deuteron properties for the OPE and TPE
potentials with spectral regularization. The computation is made by 
fixing $\gamma$ and $\eta$ to their experimental values or by fixing
$\gamma$ and predicting $\eta$ depending on the singularity structure
of the potential and the coordinate space cut-off. 
For each case we present three computations: (i) the complete computation,
i.e. the results obtained when the cut-off is completely removed and
there is no spectral cut-off, (ii) the spectrally regularized computation
with $r_c = 0.1\,{\rm fm}$, and (iii) the spectrally regularized computation 
with $r_c = 0.5\,{\rm fm}$. 
The errors quoted in the (spectrally) unregularized TPE computations and
in the spectrally regularized potential with a finite cut-off 
$r_c = 0.5\,{\rm fm}$ reflect the uncertainty in the 
non-potential parameters $\gamma$ and $\eta$. 
In the spectrally regularized potentials, the errors represent the cut-off 
dependence of the results for $r_c$ ranging from $0.1$ to $0.2\,{\rm fm}$. 
For the OPE (LO) we take 
$g_{\pi NN}=13.1(1)$. We take set IV~\cite{Entem:2003ft} for the LEC's in the
TPE calculation. For the $\Delta$ case we use Fits 1 and 2 of
Ref.~\cite{Krebs:2007rh}. Fit 1 involves the SU(4) quark-model relation
$h_A= 3 g_A /(2 \sqrt{2}) = 1.34 $ for $g_A=1.26$. Fit 2 takes 
$h_A = 1.05$.}
\end{table*}

\subsection{Remarks on Spectral Regularization}

The calculation of the NN potentials can advantageously be carried out
by using the method of dispersion
relations~\cite{Kaiser:1997mw,Kaiser:1998wa,Krebs:2007rh}.
This has motivated the use of the so-called spectral
regularization~\cite{Epelbaum:2003gr,Epelbaum:2003xx,Epelbaum:2004fk,
Krebs:2007rh}
in NN calculations where finite momentum space cut-offs have been
implemented. Remarkably, a good fit to the $^1S_0$ phase was achieved.
This is in contrast to the N3LO computation carried out in
Ref.~\cite{Entem:2007jg} with one counterterm, where a systematic
underestimation of the data was found. In our view, the relevant issue
is to disentangle the cut-off artifacts from clearly attributable
physical effects. In this section we analyze the issue of cutting-off
the potential.

The two pion exchange potential satisfies a dispersion relation based
on the $N\bar N \to 2 \pi $ amplitude. One has the representations
\begin{eqnarray}
V_C (r, {\tilde \Lambda}) &=& \frac{1}{2\pi^2 r} \int_{2 m}^{\tilde \Lambda} d\mu \mu \rho_C
(\mu) e^{-\mu r}\,, \nonumber \\ V_S (r, {\tilde \Lambda}) &=& - \frac{1}{6\pi^2 r} \int_{2
m}^{\tilde \Lambda} d\mu \mu \left(\mu^2 \rho_T(\mu) - 3 \rho_S (\mu )\right) e^{-\mu r}\,, \nonumber \\ 
V_T (r, {\tilde \Lambda}) &=& -
\frac{1}{6\pi^2 r^3} \int_{2 m}^{\tilde \Lambda} d\mu \mu^3 ( 3 + 3 \mu r +
\mu^2 r^2 ) \rho_T (\mu) e^{-\mu r} \, ,\nonumber \\   
\end{eqnarray} 
with $\rho_i (\mu) = {\rm Im} V_i ( i \mu)$ and similar relations for
the $W_i$ potentials. 
It is straightforward to see that for small $\tilde \Lambda$ and 
$r \ll {\tilde{\Lambda}}^{-1}$, the potentials behave as
\begin{eqnarray}
V_C (r, {\tilde \Lambda}) &\to & \frac{1}{2\pi^2 r} 
\int_{2 m}^{\tilde \Lambda} d\mu
\mu \rho_C (\mu) \, , \nonumber \\ V_S (r, {\tilde \Lambda}) 
&\to & - \frac{1}{6\pi^2 r} \int_{2 m}^{\tilde \Lambda} d\mu \mu 
\left( \mu^2 \rho_T (\mu ) -  3 \rho_S (\mu) \right) 
\, , \nonumber \\ 
V_T (r, {\tilde \Lambda}) &\to& - \frac{1}{6\pi^2 r^3} 
\int_{2 m}^{\tilde \Lambda} d\mu \mu^3\,3
\rho_T (\mu) \, . \nonumber \\
\end{eqnarray} 
Thus, although $V_C$, $W_C$, $V_S$ and $W_S$ become regular, the tensor
contributions $V_T$ and $W_T$ remain singular, despite higher energy
states being cut off. That means that regularization of the scattering
problem in triplet channels is mandatory to obtain well defined
results.

\begin{figure}[tbc]
\begin{center}
\epsfig{figure=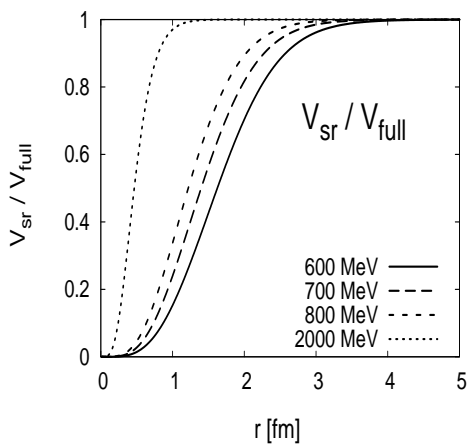,height=6cm, width=6.5cm}
\end{center}
\caption{$V(r,\Lambda) / V(r) $-ratio of $^1S_0$ potentials as a
function of the distance $r$ (in fm) for the spectral-regularized
N2LO-$\Delta$ potential with different values of the spectral cut-off
$\tilde \Lambda$ as a function of the short distance cut-off $r_c$ (in
fm).}
\label{fig:potential-spectral}
\end{figure}

\begin{figure*}[tbc]
\begin{center}
\epsfig{figure=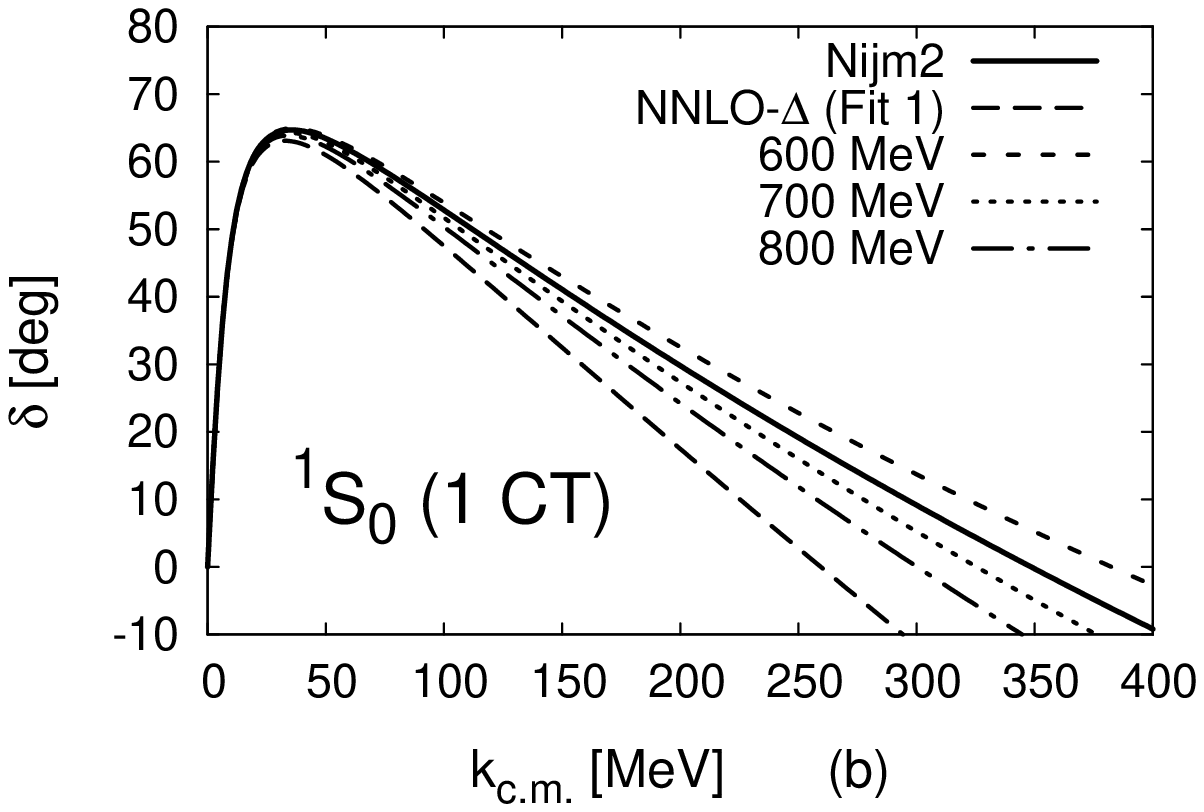,height=6cm, width=6.5cm}
\epsfig{figure=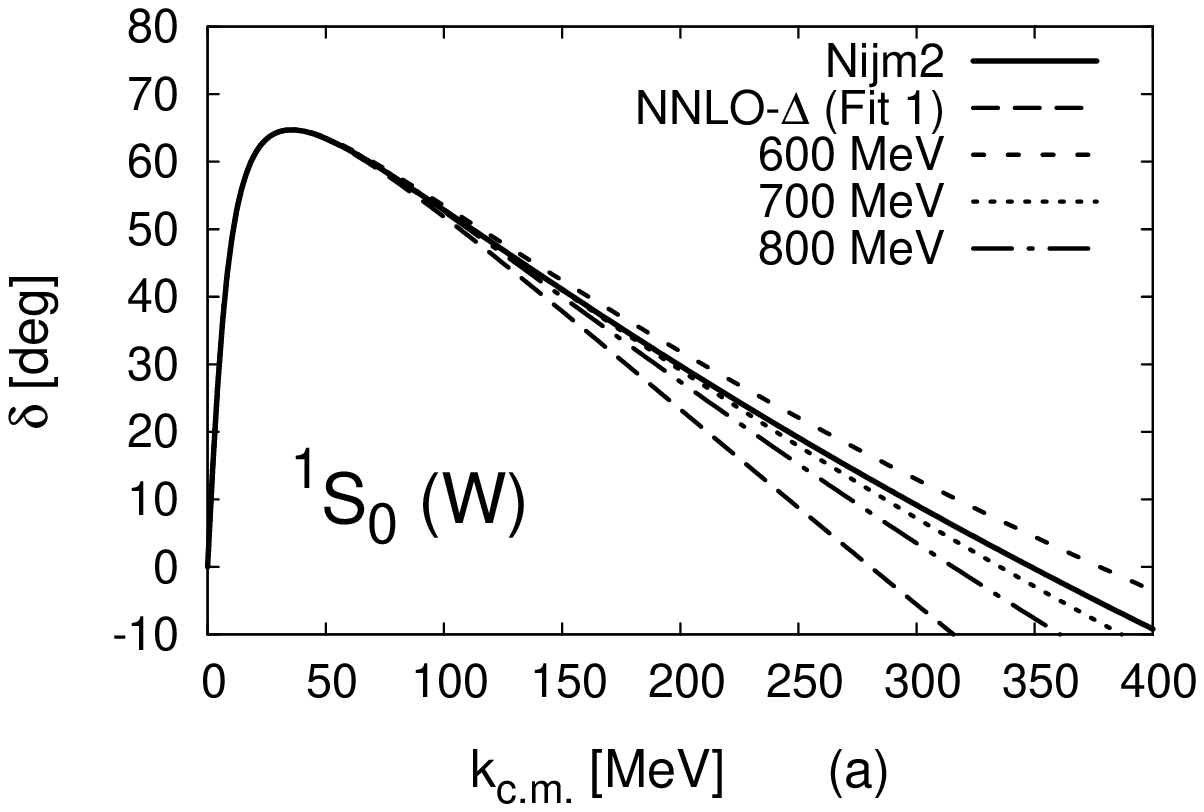, height=6cm, width=6.5cm}
\end{center}
\caption{np $^1S_0$ renormalized phase shifts for the
spectral-regularized N2LO-$\Delta$ potential for different values of
the spectral cut-off $\tilde \Lambda$ as a function of the c.m. momentum
(in MeV) compared to the Nijmegen II potential results~\cite{Stoks:1994wp} 
with one counter-term (left panel) and with two counter-terms (right panel),
corresponding this last choice to the standard Weinberg counting.}
\label{fig:ps-pcm-spectral}
\end{figure*}

Note that this short distance behaviour is quite different from the
one obtained for $r > 0 $ when ${\tilde \Lambda} \to \infty$ (see
appendix~\ref{appendix}). In Fig.~\ref{fig:potential-spectral} the
ratio, $V(r,{\tilde \Lambda}) / V(r) $, of $^1S_0$ potentials as they enter in
the phase shift calculation with and without spectral regularization
is depicted.  We note that there is a sizeable distortion at 2-3 fm's
with $\tilde \Lambda = 700\,{\rm MeV}$, decreasing the strength of the
interaction and hence providing effectively a repulsion in the singlet
potential. The effects of the spectral regularization on the potential
can be seen in Fig.~\ref{fig:ps-pcm-spectral} for NLO-$\Delta$ and
N2LO-$\Delta$ both with one counter-term as well as with two
counter-terms. As we see, it is possible to describe the data
successfully for suitable values of the spectral cut-off $\tilde
\Lambda\sim 700\,{\rm MeV}$, in agreement with the findings of 
Ref.~\cite{Krebs:2007rh}.
It is noteworthy that this happens {\it
regardless} on the usage of one or two counter-terms, reinforcing the
conclusion that the agreement is mainly due to the distortion in the
potential.
In regard to the behaviour with respect to spectral regularization,
the trends presented for the $\Delta$-full theory are also
reproduced in the $\Delta$-less scheme.
We remind that an accurate
description of the data was achieved in Ref.~\cite{Epelbaum:2004fk} at
N3LO-$\Delta \negate$ with a spectral regularized potential with
finite momentum space cut-offs and 4 counter-terms in the $^1S_0$ 
channel (the full N3LO computation has 24 counter-terms). 
On the other hand the data were
not described for large values of $p$ in the N3LO-$\Delta \negate$ and
(spectrally) unregularized renormalized calculation of
Ref.~\cite{Entem:2007jg}.

As before, it is interesting to analyze the relevant scales building
up the effective range in the case of renormalization with one counter-term.
In Fig.~\ref{fig:r0-rc-spectral}, we show the effective range $r_0$ 
as a function of the short distance cut-off $r_c$ for several values 
of the spectral cut-off $\tilde \Lambda$. 
Clearly, the stability region in $r_c$ is shifted towards
lower values as the spectral cut-off is decreased. This is consistent
with the large distortion of the potential at intermediate scales. 

We have also analyzed the impact of the spectral regularization on the
already successful description of the deuteron presented in previous 
papers and the present one.
The results do not change noticeably after (spectrally) regularizing the
potential: they lie between the results obtained with the OPE 
potential and with NLO- and N2LO-$\Delta$. This is compatible with
the weakening of these contributions to the chiral potential.
Although the final results are rather simple to summarize,
some remarks should be added on the renormalization procedure for
the deuteron channel when spectral regularization is applied. 
The modified NLO- and N2LO-$\Delta$ chiral potentials have now an attractive 
and repulsive eigen-channel, instead two attractive ones, which lead
us to the following alternative: either we remove the cutoff, in which
case we can only fix the binding energy and obtain the D-to-S ratio $\eta$
as a prediction, or we keep a finite cutoff, and then fix both the
binding energy and $\eta$. Provided that the finite cut-off is sensible,
about $r_c \sim 0.5 \,{\rm fm}$ (i.e. the saturation scale) for a spectral 
cutoff $\tilde \Lambda \sim 600-800\,{\rm MeV}$, 
both procedures give equivalent results.
The previous discussion on the effects of spectral cut-off agree with the 
remarks presented in Ref.~\cite{Valderrama:2007ja} regarding the effects
of this regularization of the potential for the deuteron form factors.
For completeness we show the results in Table~\ref{tab:deut_prop_sr}.

In summary, reducing the strength of the potential at short distances 
by means of the spectral regularization is phenomenologically preferred 
in the singlet
channel and innocuous in the triplet channel, but distorts largely the
chiral potential in a region much larger than the nucleon size. 
Moreover, we find that within our scheme the agreement with the
data is achieved {\it regardless} of the additional counter-terms
invoked by Weinberg's counting, being in fact superfluous after
renormalization. 
This said, although the spectral regularization improves over the standard
finite cut-off approaches and seems phenomenologically favoured,
the inclusion of finite size effects in a model independent manner 
would certainly be very useful.

\section{Conclusions} 
\label{sec:conclusions} 

Since more than fifteen years there has been a growing interest in
pursuing an EFT description of NN scattering below pion production
threshold where the spontaneous breakdown of chiral symmetry in QCD 
is manifestly exploited. One of the reasons which have greatly hindered
the EFT developments within the NN context has been the lack of a
credible power counting for the potential which at the same time
complies to short distance insensitivity. Given the tight constraints
under which this might actually happen, it has not been obvious which
particular form of the chiral expansion indeed embodies these
desirable properties. In the present work we have shown how the
inclusion of $\Delta$ degrees of freedom in the potential not only
complies to a phenomenologically well founded motivation, but also
provides the requested short distance insensitivity of the central
phases and the deuteron properties after the necessary renormalization
is carried out. This improves the previous situation without explicit
$\Delta$ degrees of freedom: while at LO-$\Delta \negate$ and
N2LO-$\Delta \negate$ the existence of a deuteron was compatible with
renormalizability, at NLO-$\Delta \negate$ that was not the case. This
has raised reasonable doubts on the suitability and usefulness of
non-perturbative renormalization {\it per se} to chiral potentials.  
A very rewarding aspect of the present investigation is the existence of
a deuteron bound state at LO, NLO-$\Delta$ and N2LO-$\Delta$. 
Of course, a proof of consistency to all orders, including relativistic,
spin-orbit, three pion exchange corrections, etc., remains to be done. 
This is so because, although the power law behaviour of the chiral
potential at short distances can be trivially fixed by dimensional
arguments and power counting, the determination of the attractive-repulsive 
character of the potential can so far only be fixed by actual calculations. 
The found deuteron properties seem to obey a converging pattern and 
in the $E_1$ phase a clear improvement is observed at N2LO-$\Delta$. 
Moreover, our N2LO-$\Delta$ results resemble much those of 
N2LO-$\Delta \negate$ after renormalization for reasonable parameter 
values describing the $\pi N$ reaction close to threshold.  
This suggests that despite the previous inconsistency in
NLO-$\Delta \negate$, the N2LO-$\Delta \negate$ deuteron wave functions
can be used for practical purposes, despite the theoretically
unpleasant and disturbing ``jumping'' of the NLO-$\Delta \negate$ calculation. 
In addition, it should be reminded that within these
approximations the $\pi N$ threshold properties can be fitted, 
with the sole exception of $b_{0,+}^{-}$, 
and thus the overall consistency between the $\pi N$ and $NN$ sectors
is almost satisfied.
Therefore, the inclusion of the $\Delta$ resonance complies with 
the original EFT motivation of describing {\it simultaneously}
$\pi N$ and $NN$ scattering at low energies.

We have also found that at the level of approximations involved in the
present and previous renormalized calculations, the $^1S_0$ phase shift
is not entirely reproduced for c.m. momenta larger than the pion mass 
if we insist on a reasonable $\pi N$ physics.
Several effects might be responsible for this persistent discrepancy.  
We have discussed those on the light of the relevant scales, 
$r \ge r_c = 0.5\,{\rm fm}$, which practically provide the total 
contribution to observables. 
We have further discussed parameterizations of the NN force based on a
spectral regularization of the NN potential with a cut-off of 
${\tilde \Lambda} = 700 \, {\rm MeV}$. 
We have shown that agreement to data is achieved mainly due to a large 
distortion of the potential at 2-3 fm and regardless on additional inclusion 
of counter-terms. These length scales are much larger than the expected finite 
nucleon size or vector meson exchange effects, casting doubts on 
the usefulness of spectral regularization and suggesting the need for a more 
controllable and  better founded description of the missing short 
distance physics in the $^1S_0$ channel.

In the present paper we have restricted to central phases and the
deuteron, but the calculation of higher partial waves is also of
interest, as well as the inclusion of Coulomb effects in pp
scattering. In all cases, the negative definite character of the
potential at short distances guarantees the existence of convergent results. 
Finally, the present results can have some impact on
calculations including deuteron properties such as deuteron form
factors, pion-deuteron scattering, and further low energy matrix
elements of electroweak deuteron reactions where short distance
insensitivity and chiral symmetry are both expected to play a 
significant role.

\begin{acknowledgments}

We thank Andreas Nogga, Evgeny Epelbaum, Ulf-G. Mei{\ss}ner and
Johann Haidenbauer for discussions and a critical and 
careful reading of the manuscript.
M. P. V. is supported
by the Helmholtz Association fund provided to the young investigator group 
``Few-Nucleon Systems in Chiral Effective Field Theory''  (grant VH-NG-222),
and to the virtual institute ``Spin and strong QCD'' (VH-VI-231).
The work of E.R.A.  is supported in part by funds provided by the Spanish
DGI and FEDER funds with grant no. FIS2005-00810, Junta de
Andaluc{\'\i}a grants no.  FQM225-05, EU Integrated Infrastructure
Initiative Hadron Physics Project contract no. RII3-CT-2004-506078.

\end{acknowledgments}

\appendix

\section{Short distance expansion of the potentials}
\label{appendix}

Using the spectral representation the short distance expansion of the
potentials can be done by expanding spectral functions for $\mu \gg
m$, with $\mu r $ fixed. This way one obtains 
\begin{eqnarray}
V_C(r) &=& \frac{C_6^{V,C}}{r^6} + \dots \nonumber \\ 
W_C(r) &=& \frac{C_6^{W,C}}{r^6} + \dots \nonumber \\ 
V_S(r) &=& \frac{C_6^{V,S}}{r^6} + \dots \nonumber \\ 
W_S(r) &=& \frac{C_6^{W,S}}{r^6} + \dots \nonumber \\ 
V_T(r) &=& \frac{C_6^{V,T}}{r^6} + \dots \nonumber \\ 
W_T(r) &=& \frac{C_6^{W,T}}{r^6} + \dots  
\end{eqnarray} 
where the van der Waals coefficients at NLO-$\Delta$ are given by
\begin{eqnarray}
C_{6,V,C}^{\rm NLO-\Delta} &=& - \frac{(9 g_A^2 + 4 h_A^2)
h_A^2}{36 f_\pi^4 \pi^2 \Delta}\\ 
C_{6,W,C}^{\rm NLO-\Delta} &=& - \frac{(9 g_A^2 - 2 h_A^2)
h_A^2}{108 f_\pi^4 \pi^2 \Delta} \\ 
C_{6,V,S}^{\rm NLO-\Delta} &=& \frac{(9 g_A^2 - 2 h_A^2)
h_A^2}{216 f_\pi^4 \pi^2 \Delta} \\
C_{6,W,S}^{\rm NLO-\Delta} &=& \frac{(9 g_A^2 + h_A^2)
h_A^2}{648 f_\pi^4 \pi^2 \Delta} \\
C_{6,V,T}^{\rm NLO-\Delta} &=& - \frac{(9 g_A^2 - 2 h_A^2)
h_A^2}{216 f_\pi^4 \pi^2 \Delta} \\
C_{6,W,T}^{\rm NLO-\Delta} &=& - \frac{(9 g_A^2 + h_A^2)
h_A^2}{648 f_\pi^4 \pi^2 \Delta} 
\end{eqnarray} 
and at N2LO-$\Delta$ by 
\begin{eqnarray}
C_{6,V,C}^{\rm N2LO-\Delta} &=&  \frac{4 \tilde{b}\,h_A^3}{9 f_\pi^4 \pi^2} 
+\frac{c_3 h_A^2}{f_\pi^4 \pi^2} +\frac{9 c_3 g_A^2}{16 f_\pi^4 \pi^2} 
\\ 
C_{6,W,C}^{\rm N2LO-\Delta} &=&  \frac{2 \tilde{b}\,h_A^3}{27 f_\pi^4 \pi^2} 
-\frac{\tilde{b}\,g_A^2 h_A}{6 f_\pi^4 \pi^2} \\ 
C_{6,V,S}^{\rm N2LO-\Delta} &=&  \frac{\tilde{b}\,g_A^2 h_A}{12 f_\pi^4 \pi^2} 
-\frac{\tilde{b}\,h_A^3}{27 f_\pi^4 \pi^2} \\ 
C_{6,W,S}^{\rm N2LO-\Delta} &=&  -\frac{\tilde{b}\,h_A^3}{162 f_\pi^4 \pi^2} 
+\frac{c_4 h_A^2}{36 f_\pi^4 \pi^2} +\frac{c_4 g_A^2}{16 f_\pi^4 \pi^2} 
\\ 
C_{6,V,T}^{\rm N2LO-\Delta} &=&  \frac{\tilde{b}\,h_A^3}{27 f_\pi^4 \pi^2} -
\frac{\tilde{b}\,g_A^2 h_A}{12 f_\pi^4 \pi^2} 
\\ 
C_{6,W,T}^{\rm N2LO-\Delta} &=&  \frac{\tilde{b}\,h_A^3}{162 f_\pi^4 \pi^2} 
-\frac{c_4 h_A^2}{36 f_\pi^4 \pi^2}
-\frac{c_4 g_A^2}{16 f_\pi^4 \pi^2} 
\end{eqnarray} 
where $\tilde{b} = b_3 + b_8$.

%\bibliography{Delta}

%\end{document}

\end{document}